\documentclass[10pt,twocolumn]{article}
\usepackage[top=1.9cm, bottom=1.9cm, left=1.7cm, right=1.7cm]{geometry}
\usepackage{times}
\usepackage[hyphens]{url}  %
\usepackage[keeplastbox]{flushend}  %
\usepackage{graphicx}  %
\usepackage{xcolor}
\usepackage{flushend}
\usepackage{booktabs}
\usepackage{graphicx}
\usepackage{paralist}
\usepackage{enumitem}
\usepackage[small,bf]{caption}
\usepackage{subfigure}
\usepackage[hang,flushmargin]{footmisc}
\renewcommand{\footnotesize}{\fontsize{8}{9}\selectfont}
\usepackage[bookmarks=true,%
           bookmarksnumbered=true,%
           colorlinks=true,%
            linkcolor=linkcol,%
            citecolor=citecol,%
            urlcolor=urlcol,%
            hypertexnames=true,%
            pdfpagelabels,
	    ]%
      {hyperref}

\usepackage[OT2,OT1]{fontenc}
\newcommand\cyr
{
\renewcommand\rmdefault{wncyr}
\renewcommand\sfdefault{wncyss}
\renewcommand\encodingdefault{OT2}
\normalfont
\selectfont
}
\DeclareTextFontCommand{\textcyr}{\cyr}

\usepackage[compact]{titlesec}
\titlespacing*{\section}{0pt}{*4}{4pt} 
\titlespacing{\subsection}{0pt}{*3}{3pt}

\definecolor{linkcol}{rgb}{0,0,0.5}
\definecolor{citecol}{rgb}{0,0.5,0.3}
\definecolor{urlcol}{rgb}{0.3,0,0}

\renewenvironment{thebibliography}[1]{
  \begin{oldthebibliography}{#1}
    \setlength{\itemsep}{0.1em}
    \setlength{\parskip}{0.1em}
}
{
  \end{oldthebibliography}
}

\renewcommand{\footnoterule}{%
  \kern -3pt
  \hrule width 1in
  \kern 2pt
}

\usepackage{xspace}

\newcommand{\descr}[1]{\smallskip\noindent\textbf{#1}}

\newcommand{\dspol}{{/pol/}\xspace}

\makeatletter
\def\url@leostyle{%
  \@ifundefined{selectfont}{\def\UrlFont{}}%
  {\def\UrlFont{}}%
}
\makeatother
\usepackage[hyphenbreaks]{breakurl}

\newif\ifwatermark
\watermarkfalse

\ifwatermark
    \usepackage{draftwatermark}
    \SetWatermarkScale{.7}
    \SetWatermarkText{\shortstack{In Submission:\\Do Not Distribute}}
    \SetWatermarkColor[rgb]{1,0.88,0.88}
\fi

\usepackage{etoolbox}
\makeatletter
\patchcmd\@combinedblfloats{\box\@outputbox}{\unvbox\@outputbox}{}{%
   \errmessage{\noexpand\@combinedblfloats could not be patched}%
}%
 \makeatother
\AtBeginShipout{%
  \ifnum\value{page}>1 %
    \typeout{* Additional boxing of page `\thepage'}%
    \setbox\AtBeginShipoutBox=\hbox{\copy\AtBeginShipoutBox}%
  \fi
}

\begin{document}

\title{\LARGE \bf Who Let The Trolls Out? \\Towards Understanding State-Sponsored Trolls}
\author{Savvas Zannettou$^{\star\ddagger}$, Tristan Caulfield$^\dagger$, William Setzer$^\ddagger$,\\Michael Sirivianos$^{\star}$, Gianluca Stringhini$^\diamond$, Jeremy Blackburn$^\ddagger$\\[0.5ex]
\normalsize $^{\star}$Cyprus University of Technology, $^\dagger$University College London, ${^\ddagger}$University of Alabama at Birmingham, ${^\diamond}$Boston University\\
\normalsize sa.zannettou@edu.cut.ac.cy, t.caulfield@ucl.ac.uk, wjsetzer@uab.edu, %
\normalsize michael.sirivianos@cut.ac.cy, blackburn@uab.edu}
\date{}

\maketitle
\begin{abstract}
Recent evidence has emerged linking coordinated campaigns by state-sponsored actors to manipulate public opinion on the Web.
Campaigns revolving around major political events are enacted via mission-focused ``trolls.''
While trolls are involved in spreading disinformation on social media, there is little understanding of how they operate, what type of content they disseminate, how their strategies evolve over time, and how they influence the Web's information ecosystem.

In this paper, we begin to address this gap by analyzing 10M posts by 5.5K Twitter and Reddit users identified as Russian and Iranian state-sponsored trolls.
We compare the behavior of each group of state-sponsored trolls with a focus on how their strategies change over time, the different campaigns they embark on, and differences between the trolls operated by Russia and Iran.
Among other things, we find: 1)~that Russian trolls were pro-Trump while Iranian trolls were anti-Trump; 
2)~evidence that campaigns undertaken by such actors are influenced by real-world events; and 
3)~that the behavior of such actors is not consistent over time, hence automated detection is not a straightforward task.
Using the Hawkes Processes statistical model, we quantify the influence these accounts have on pushing URLs on four social platforms: Twitter, Reddit, 4chan's Politically Incorrect board (\dspol), and Gab.
In general, Russian trolls were more influential and efficient in pushing URLs to all the other platforms with the exception of \dspol where Iranians were more influential.
Finally, we release our data and source code to ensure the reproducibility of our results and to encourage other researchers to work on understanding other emerging kinds of state-sponsored troll accounts on Twitter.

\end{abstract}

\section{Introduction}

Recent political events and elections have been increasingly accompanied by reports of disinformation campaigns attributed to state-sponsored actors~\cite{ferrara2017disinformation}.
In particular, ``troll farms,'' allegedly employed by Russian state agencies, have been actively commenting and posting content on social media to further  the Kremlin's political agenda~\cite{independent}.

Despite the growing relevance of state-sponsored disinformation, the activity of accounts linked to such efforts has not been thoroughly studied.
Previous work has mostly looked at campaigns run by bots~\cite{ferrara2017disinformation,hegelich2016are,ratkiewicz2011detecting}.
However, automated content diffusion is only a part of the issue. In fact, recent research has shown that human actors are actually key in spreading false information on Twitter~\cite{starbird2017examining}.
Overall, many aspects of state-sponsored disinformation remain unclear, e.g., how do state-sponsored trolls operate? What kind of content do they disseminate? How does their behavior change over time? And, perhaps more importantly, is it possible to quantify the influence they have on the overall information ecosystem on the Web?

In this paper, we aim to address these questions, by relying on two different sources of ground truth data about state-sponsored actors. First, we use 10M tweets posted by Russian and Iranian trolls between 2012 and 2018~\cite{twitter_russian_iranians_dataset}.
Second, we use a list of 944 Russian trolls, identified by Reddit, and find all their posts between 2015 and 2018~\cite{reddit_dataset_trolls}.
We analyze the two datasets across several axes in order to understand their behavior and how it changes over time, their targets, and the content they shared.
For the latter, we leverage word embeddings to understand in what context specific words/hashtags are used and shed light to the ideology of the trolls.
Also, we use Hawkes Processes~\cite{linderman2014} to model the influence that the Russian and Iranian trolls had over multiple Web communities; namely, Twitter, Reddit, 4chan's Politically Incorrect board (\dspol)~\cite{hine2016longitudinal}, and Gab~\cite{zannettou2018gab}.

\descr{Main findings.}
Our study leads to several key observations:
\begin{enumerate}%
\item Our influence estimation experiments reveal that Russian trolls were extremely influential and efficient in spreading URLs on Twitter. Also, when we compare their influence and efficiency to Iranian trolls, we find that Russian trolls were more efficient and influential in spreading URLs on Twitter, Reddit, Gab, but not on \dspol.
\item By leveraging word embeddings, we find ideological differences between Russian and Iranian trolls. For instance, we find that Russian trolls were pro-Trump, while Iranian trolls were anti-Trump.
\item We find evidence that the Iranian campaigns were motivated by real-world events. Specifically, campaigns against France and Saudi Arabia coincided with real-world events that affect the relations between these countries and Iran.
\item We observe that the behavior of trolls varies over time. We find substantial changes in the use of language and Twitter clients over time for both Russian and Iranian trolls. These insights allow us to understand the targets of the orchestrated campaigns for each type of trolls over time.
\item We find that the topics of interest and discussion vary across Web communities. For example, we find evidence that Russian trolls on Reddit were extensively discussing about cryptocurrencies, while this does not apply in great extent for the Russian trolls on Twitter.
\end{enumerate}

Finally, we make our source code publicly available~\cite{code} for reproducibility purposes and to encourage researchers to further work on understanding other types of state-sponsored trolls on Twitter (i.e., on January 31, 2019, Twitter released data related to state-sponsored trolls originating from Venezuela and Bangladesh~\cite{new_twitter_dataset}).

\section{Related Work} \label{sec:related}
We now review previous work on opinion manipulation as well as politically motivated disinformation on the Web. %

\descr{Opinion manipulation.}
The practice of swaying opinion in Web communities has become a hot-button issue as malicious actors are intensifying their efforts 
to push their subversive agenda.
Kumar et al.~\cite{kumar2017sockpuppets} study how users create multiple accounts, called \emph{sockpuppets}, that actively participate in some communities with the goal to manipulate users' opinions.
Mihaylov et al.~\cite{mihaylov2015finding} show that trolls can indeed manipulate users' opinions in online forums.
In follow-up work, Mihaylov and Nakov~\cite{mihaylov2016hunting} highlight two types of trolls: those paid to operate and those that are called out as such by other users.
Then, Volkova and Bell~\cite{volkova2016account} aim to predict the deletion of Twitter accounts because they are trolls,
focusing on those that shared content related to the Russia-Ukraine crisis.

Elyashar et al.~\cite{elyashar2017is} distinguish authentic discussions from campaigns to manipulate the public's opinion,
using a set of similarity functions alongside historical data. %
Also, Steward et al.~\cite{steward2018examining} focus on discussions related to the Black Lives Matter movement and how content from Russian trolls was retweeted by other users.
Using community detection techniques, they unveil that Russian trolls infiltrated both left and right leaning communities, setting out to push specific narratives.
Finally, Varol et al.~\cite{varol2017early} aim to identify memes (ideas) that become popular due to \emph{coordinated} efforts,
and achieve %
a 75\% AUC score before memes become trending and a 95\% AUC score afterwards.

\descr{False information on the political stage.}
Conover et al.~\cite{conover2011political} focus on Twitter activity over the six weeks leading to the 2010 US midterm elections and the interactions between right and left leaning communities.
Ratkiewicz et al.~\cite{ratkiewicz2011detecting} study political campaigns using multiple controlled accounts to disseminate support for an individual or opinion. Specifically, they use machine learning to detect the early stages of false political information spreading on Twitter.
Wong et al.~\cite{wong2013quantifying} aim to quantify the political leanings of users and news outlets during the 2012 US presidential election on Twitter by formulating the problem as an ill-posed linear inverse problem, and using an inference engine that considers tweeting and retweeting behavior of articles.
Yang et al.~\cite{yang2016social} investigate the topics of discussions on Twitter for 51 US political persons
showing that Democrats and Republicans are active in a similar way on Twitter, although the former tend to use hashtags more frequently.
Le et al.~\cite{le2017revisiting} study 50M tweets pertaining to the 2016 US election primaries and highlight the importance of three factors in political discussions on social media, namely the \textit{party} (e.g., Republican or Democrat), \textit{policy considerations} (e.g., foreign policy), and \textit{personality} of the candidates (e.g., intelligent or determined).

Howard and Kollanyi~\cite{howard2016bots} study the role of bots in Twitter conversations during  the 2016 Brexit referendum.
They find that most tweets are in favor of Brexit, that there are bots with various levels of automation, and that 1\% of the accounts generate 33\% of the overall messages.
Also, Hegelich and Janetzko~\cite{hegelich2016are} investigate whether bots on Twitter are used as political actors.
By exposing and analyzing 1.7K bots on Twitter during the Russia-Ukraine conflict, they uncover their political agenda and show that bots exhibit various behaviors, e.g., trying to hide their identity, promoting topics through the use of hashtags, and retweeting messages with particularly interesting content.
Badawy et al.~\cite{badawy2018falls} aim to predict users that are likely to spread information from state-sponsored actors, while Dutt et al.~\cite{dutt2018senator} focus on the Facebook platform and analyze ads shared by Russian trolls in order to find the cues that make them effective.
Finally, a large body of work focuses on social bots~\cite{bessi2016social,davis2016bot,ferrara2016the,ferrara2017disinformation,varol2017online} and their role in spreading political disinformation, highlighting that they can manipulate the public's opinion at a large scale, thus potentially affecting the outcome of political elections.

\descr{Remarks.} Unlike previous work, our study focuses on a set of Russian and Iranian trolls that were suspended by Twitter and Reddit.
To the best of our knowledge, this constitutes the first effort not only to characterize a ground truth of troll accounts independently identified by Twitter and Reddit, but also to quantify their influence on the greater Web, specifically, on Twitter as well as on other communities like Reddit, 4chan, and Gab.

\section{Background} \label{sec:background}
In this section, we provide a brief overview of the social networks studied in this paper, i.e., Twitter, Reddit, 4chan, and Gab,
which we choose because they are impactful actors on the Web's information ecosystem~\cite{zannettou2017web,zannettou2018origins,zannettou2018gab,hine2016longitudinal}.
Note that the two latter Web communities are only used in our influence estimation experiments (see Section~\ref{sec:influence}), where we aim to understand the influence that trolls had to these Web communities.

\descr{Twitter.} Twitter is a mainstream social network, where users can broadcast short messages, called ``tweets,'' to their followers.
Tweets may contain hashtags, which enable the easy index and search of messages, as well as mentions, which refer to other users on Twitter.

\descr{Reddit.} Reddit is a news aggregator with several social features.
It allows users to post URLs along with a title; posts can get up-  and down- votes, which dictate the popularity and order in which they appear on the platform.
Reddit is divided to ``subreddits,'' which are forums created by users that focus on a particular topic (e.g., /r/The\_Donald is about discussions around Donald Trump).

\descr{4chan.} 4chan is an imageboard  forum, organized in communities called ``boards,'' each with a different topic of interest.
A user can create a new post by uploading an image with or without some text; others can reply below with or without images.
4chan is an anonymous community, and several of its boards are reportedly responsible for a substantial amount of hateful content~\cite{hine2016longitudinal}.
In this work we focus on the Politically Incorrect board (\dspol) mainly because it is the main board for the discussion of politics and world events.
Furthermore, 4chan is ephemeral, i.e., there is a limited number of active threads and all threads are permanently deleted after a week.
We collect our 4chan dataset, between June 30, 2016, and October 20, 2018, using the methodology described in~\cite{hine2016longitudinal}, ultimately collecting 98M posts.

\descr{Gab.} Gab is a social network launched in August 2016 aiming to provide a platform for free speech and explicitly welcomes users banned from other communities..
It combines features from Twitter (broadcast of 300-character messages, called ``gabs'') and Reddit (content popularity according to a voting system).
It also has extremely lax moderation policies; it allows everything except illegal pornography, terrorist propaganda, and doxing~\cite{snyder2017fifteen}.
Overall, Gab attracts alt-right users, conspiracy theorists, and high volumes of hate speech~\cite{zannettou2018gab}.
We collect 46M posts, posted on Gab between August 10, 2016 and October 20, 2018, using the same methodology as in~\cite{zannettou2018gab}.

\section{Troll Datasets} \label{sec:datasets}

In this section, we describe our dataset of Russian and Iranian trolls on Twitter and Reddit.

\begin{table}[]
\centering
\resizebox{\columnwidth}{!}{%
\begin{tabular}{llrrr}
\hline
\multicolumn{1}{c}{\textbf{Platform}} & \multicolumn{1}{c}{\textbf{Origin of trolls}} & \multicolumn{1}{c}{\textbf{\# trolls}} & \multicolumn{1}{c}{\textbf{\begin{tabular}[c]{@{}c@{}}\# trolls\\ with tweets/posts\end{tabular}}} & \multicolumn{1}{l}{\textbf{\# of tweets/posts}} \\ \hline
\textbf{Twitter} & \textbf{Russia} & 3,836 & 3,667 & 9,041,308 \\
 & \textbf{Iran} & 770 & 660 & 1,122,936 \\ \hline
\textbf{Reddit} & \textbf{Russia} & 944 & 335 & 21,321 \\ \hline
\end{tabular}
}
\caption{Overview of Russian and Iranian trolls on Twitter and Reddit. We report the overall number of identified trolls, the trolls that had at least one tweet/post, and the overall number of tweets/posts.}
\label{tbl:trolls_datasets}
\end{table}

\descr{Twitter.} On October 17, 2018, Twitter released a large dataset of Russian and Iranian troll accounts~\cite{twitter_russian_iranians_dataset}.
Although the exact methodology used to determine that these accounts were
state-sponsored trolls is unknown, based on the most recent Department of Justice indictment~\cite{trollsindictment},
the dataset appears to have been constructed in a manner that we can assume essentially no false positives, while we cannot make any postulation about false negatives.
Table~\ref{tbl:trolls_datasets} summarizes the troll dataset.

\descr{Reddit.} On April 10, 2018, Reddit released a list of 944 accounts which they determined were operated by actors working on behalf of the Russian government~\cite{reddit_dataset_trolls}.
We recover the submissions, comments, and account details for these accounts using two mechanisms:
1)~dumps of Reddit provided by Pushshift~\cite{pushshift}; and 2)~crawling the user pages of those accounts. %
Although omitted for lack of space, we note that the union of these two data sources reveals some gaps in both, likely due to a combination of subreddit moderators removing posts or the troll users themselves deleting them, which would affect the two data sources in different ways.
In any case, for our purposes, we merge the two datasets, with Table~\ref{tbl:trolls_datasets} describing the final dataset.
Note that only about one third (335) of the accounts released by Reddit had at least one submission or comment in our dataset.
We suspect the rest were simply used as dedicated upvote/downvote accounts used in an effort to push (or bury) specific content.

\descr{Ethics.} Although we only work with publicly available data, we follow standard ethical guidelines~\cite{rivers2014ethical} and make no attempt to de-anonymize users. %

\section{Analysis} \label{sec:analysis}
In this section, we present an in-depth analysis of the activities and the behavior of Russian and Iranian trolls on Twitter and Reddit.

\subsection{Accounts Characteristics}

\begin{figure}[t!]
\centering
\includegraphics[width=\columnwidth]{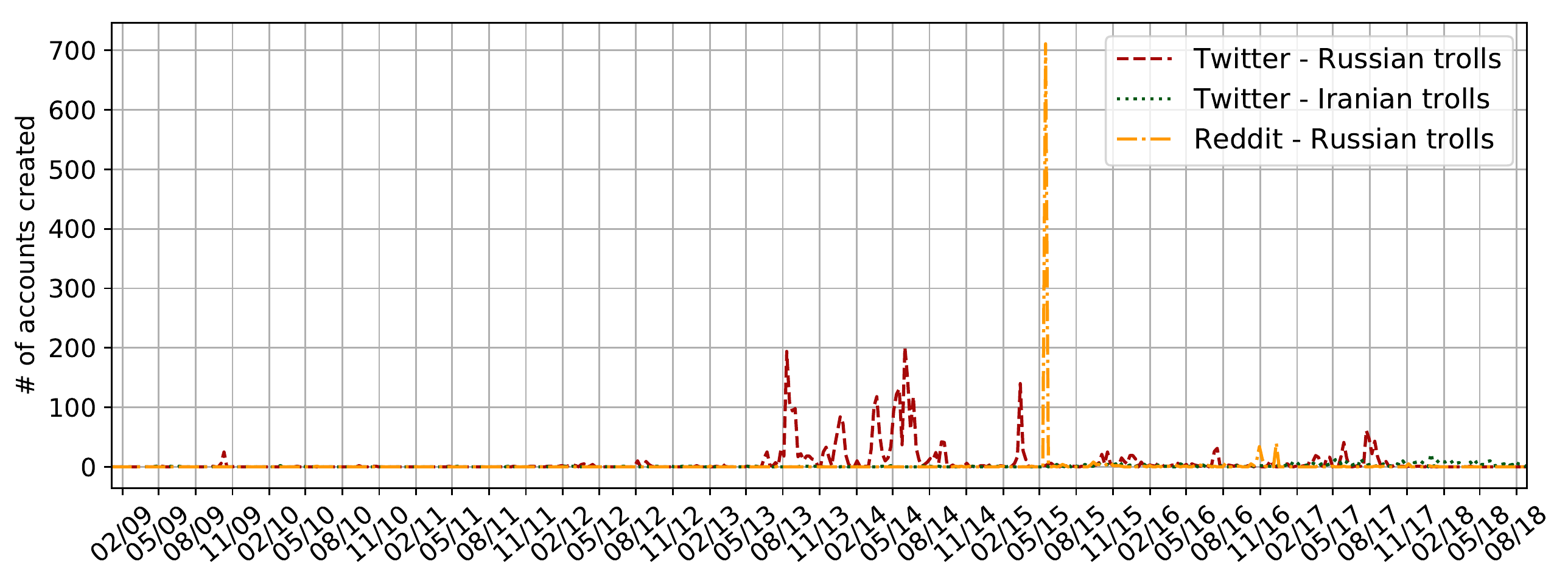}
\caption{Number of Russian and Iranian troll accounts created per week.}
\label{fig:counts_created}
\end{figure}

\begin{table}[t!]
  \centering
  \resizebox{\columnwidth}{!}{
  \setlength{\tabcolsep}{0.4em} %
\begin{tabular}{rrrrrlrl}
\hline
\multicolumn{4}{c}{\textbf{Russian troll on Twitter}} & \multicolumn{4}{c}{\textbf{Iranian trolls on Twitter}} \\ \hline
\textbf{Word} & \multicolumn{1}{l}{\textbf{(\%)}} & \textbf{Bigrams} & \multicolumn{1}{l}{\textbf{(\%)}} & \textbf{Word} & \textbf{(\%)} & \textbf{Bigrams} & \textbf{(\%)} \\ \hline
follow & 7.7\% & follow me & \multicolumn{1}{r|}{6.4\%} & journalist & 3.6\% & human rights & 1.6\% \\
love & 4.8\% & breaking news & \multicolumn{1}{r|}{0.8\%} & news & 3.2\% & independent news & 1.4\% \\
life & 4.5\% & donald trump & \multicolumn{1}{r|}{0.7\%} & independent & 2.8\% & news media & 1.4\% \\
trump & 4.4\% & lokale nachrichten & \multicolumn{1}{r|}{0.6\%} & lover (in Farsi) & 2.6\% & media organization & 1.4\% \\
conservative & 4.3\% & nachrichten aus & \multicolumn{1}{r|}{0.6\%} & social & 2.6\% & organization aim & 1.4\% \\
news & 3.4\% & hier kannst & \multicolumn{1}{r|}{0.6\%} & politics & 2.6\% & aim inspire & 1.4\% \\
maga & 3.4\% & kannst du & \multicolumn{1}{r|}{0.6\%} &  media & 2.4\% &inspire action & 1.4\% \\
{\cyr lyublyu} & 2.4\% & du wichtige & \multicolumn{1}{r|}{0.6\%} & love & 2.2\% & action likes & 1.4\% \\
will & 2.4\% & wichtige und & \multicolumn{1}{r|}{0.6\%} &  justice& 2.0\% & likes social & 1.4\% \\
proud & 2.2\% & und aktuelle & \multicolumn{1}{r|}{0.6\%} &  low (in Farsi)& 2.0\% & social justice & 1.4\% \\ \hline
\end{tabular}
  }
  \caption{Top 10 words and bigrams found in the descriptions of Russian and Iranian trolls on Twitter.}
  \label{tbl:account_desc}
\end{table}

First we explore when the accounts appeared, what they posed as, and how many followers/friends they had on Twitter.

\descr{Account Creation.} Fig.~\ref{fig:counts_created} plots the Russian and Iranian troll accounts creation dates on Twitter and Reddit.
We observe that the majority of Russian troll accounts were created around the time of the Ukrainian conflict: 80\% of have an account creation date earlier than 2016.
That said, there are some meaningful peaks in account creation during 2016 and 2017.
57 accounts were created between July 3-17, 2016, which was right before the start of the Republican National Convention (July 18-21) where Donald Trump was named the Republican nominee for President~\cite{rnc_meeting} .
Later, 190 accounts were created between July, 2017 and August, 2017, during the run up to the infamous Unite the Right rally in Charlottesville~\cite{charlotesville}.
Taken together, this might be evidence of coordinated activities aimed at manipulating users' opinions on Twitter with respect to specific events.
This is further evidenced when examining the Russian trolls on Reddit: 75\% of Russian troll accounts on Reddit were created in a single massive burst in the first half of 2015.
Also, there are a few smaller spikes occurring just prior to the 2016 US Presidential election.
For the Iranian trolls on Twitter we observe that they are much ``younger,'' with the larger bursts of account creation \emph{after} the 2016 US Presidential election.

\descr{Account Information.}
To avoid being obvious, state sponsored trolls might attempt to present a persona that masks their true nature or otherwise ingratiates themselves to their target audience.
By examining the profile description of trolls we can get a feeling for how they might have cultivated this persona.
In Table~\ref{tbl:account_desc}, we report the top ten words and bigrams that appear in profile descriptions of trolls on Twitter.
Note that we do this only for Twitter trolls as we do not have descriptions for Reddit accounts.
From the table we see that a relatively large number of Russian trolls pose as news outlets, with ``news'' (1.3\%) and ``breaking news'' (0.8\%) appearing in their description.
Further, they seem to use their profile description to more explicitly increase their reach on Twitter, by nudging users to follow them (e.g., ``follow me'' appearing in almost 6.4\% of profile descriptions).
Finally, 3.4\% of the Russian trolls describe themselves as Trump supporters: see ``trump'' (4.4\%) and ``maga'' (3.4\%) terms.
Iranian trolls are even more likely to pose as news outlets or journalists: 3.6\% have ``journalist'' and 3.2\% have ``news'' in their profile descriptions.
This highlights that accounts that pose as news outlets may in fact be accounts controlled by state-sponsored actors, hence regular users should critically think in order to assess whether the account is credible or not.

\begin{figure}[t]
\center
\subfigure[]{\includegraphics[width=0.49\columnwidth]{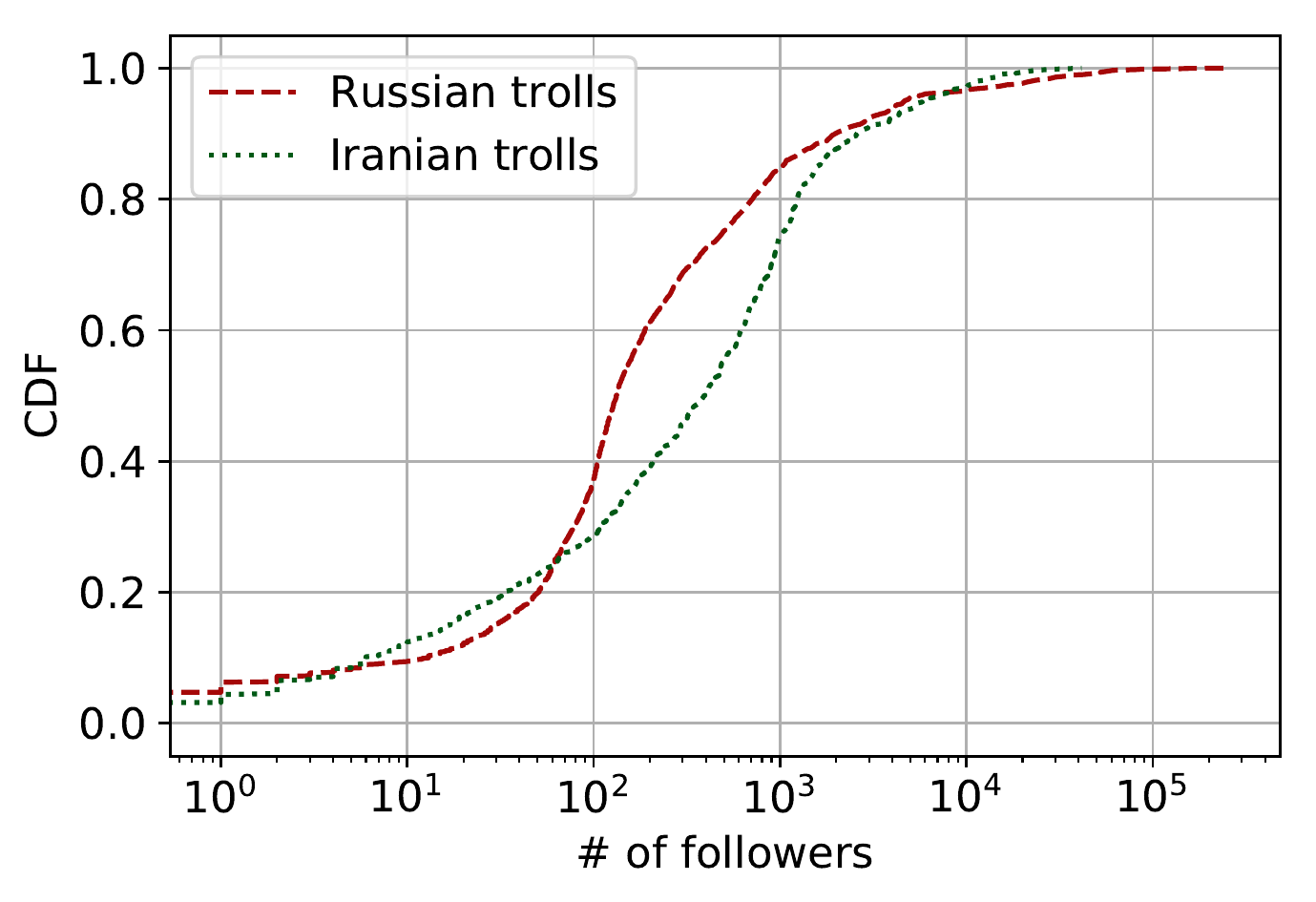}\label{subfig:cdf_followers}}
\subfigure[]{\includegraphics[width=0.49\columnwidth]{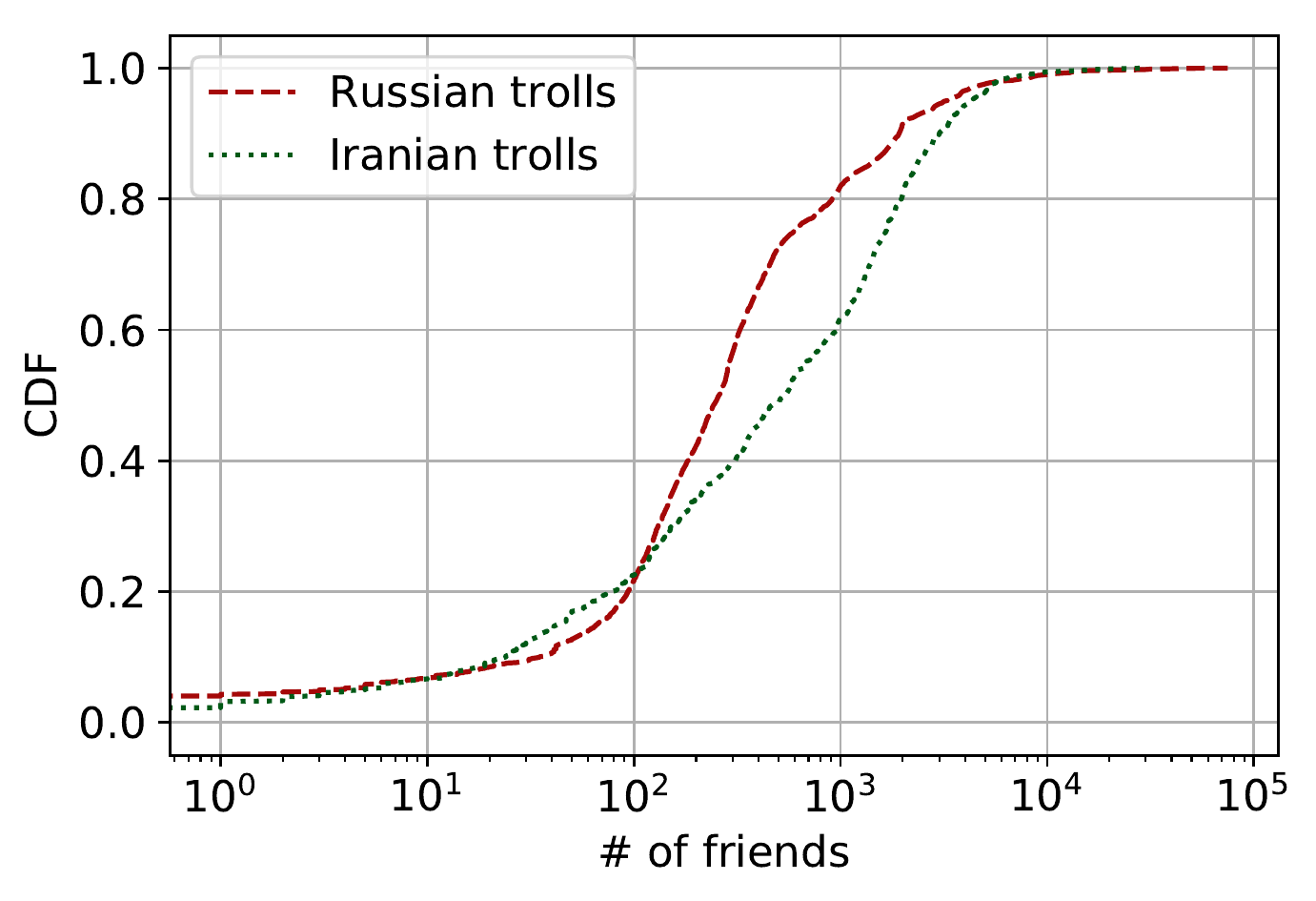}\label{subfig:cdf_followings}}
\caption{CDF of the number of a) followers and b) friends for the Russian and Iranian trolls on Twitter.}
\label{fig:cdf_followers_followings}
\end{figure}

\descr{Followers/Friends.} Fig.~\ref{fig:cdf_followers_followings} plots the CDF of the number of followers and friends for both Russian and Iranian trolls.
25\% of Iranian trolls had more than 1k followers, while the same applies for only 15\% of the Russian trolls.
In general, Iranian trolls tend to have more followers than Russian trolls (median of 392 and 132, respectively).
Both Russian and Iranian trolls tend to follow a large number of users, probably in an attempt to increase their follower count via reciprocal follows.
Iranian trolls have a median followers to friends ratio of 0.51, while Russian trolls have a ratio of 0.74.
This might indicate that Iranian trolls were more effective in acquiring followers without resorting in massive followings of other users, or perhaps that they took advantages of services that offer followers for sale~\cite{stringhini2013follow}.

\subsection{Temporal Analysis}

\begin{figure}[t]
\center
\subfigure[Date]{\includegraphics[width=0.97\columnwidth]{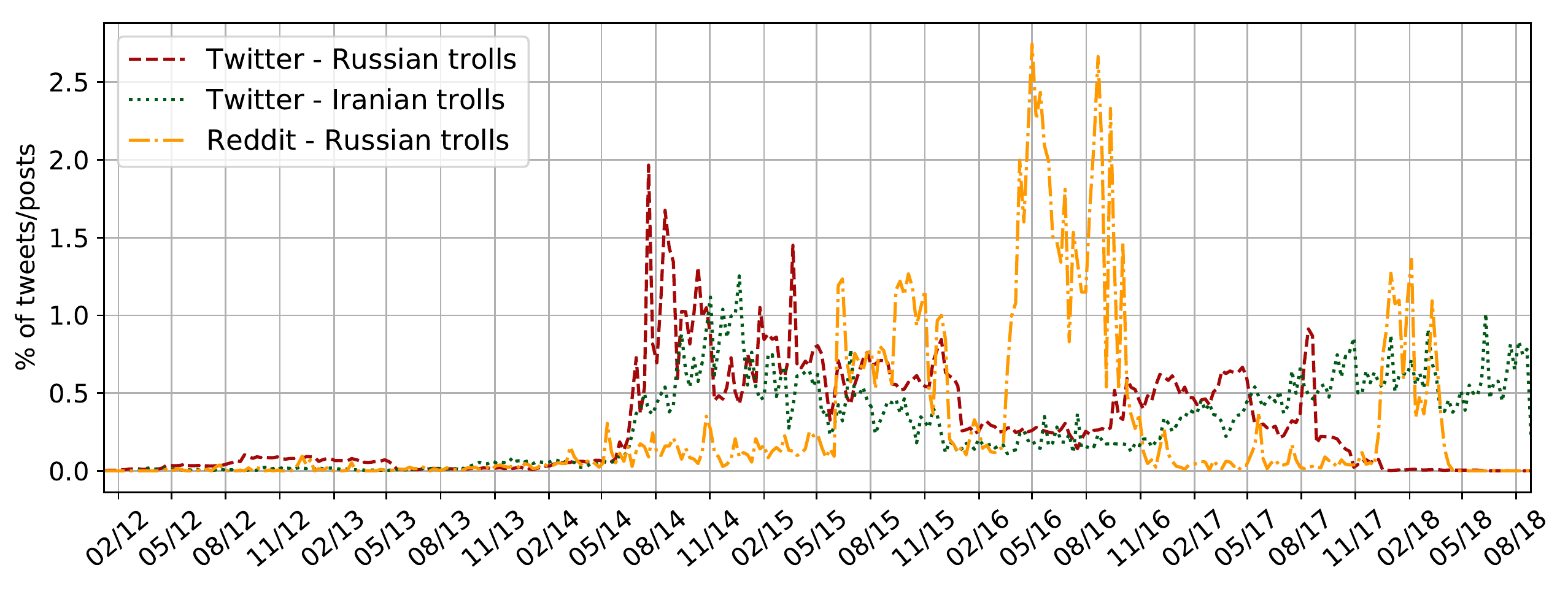}\label{subfig:counts_per_day}}
\subfigure[Hour of Day]{\includegraphics[width=0.485\columnwidth,height=1.1in]{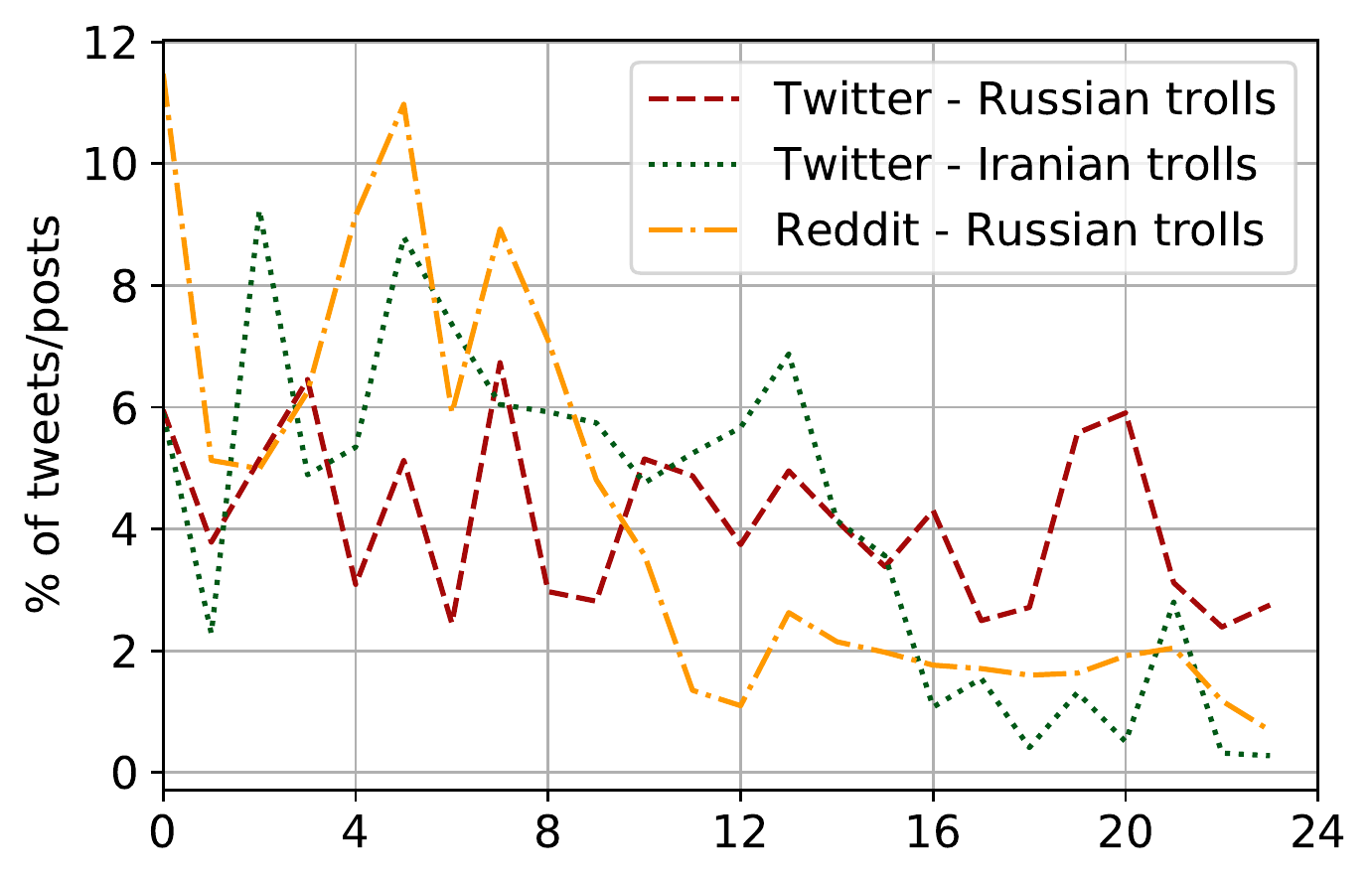}\label{subfig:counts_per_hour_day}}
\subfigure[Hour of Week]{\includegraphics[width=0.485\columnwidth,height=1.1in]{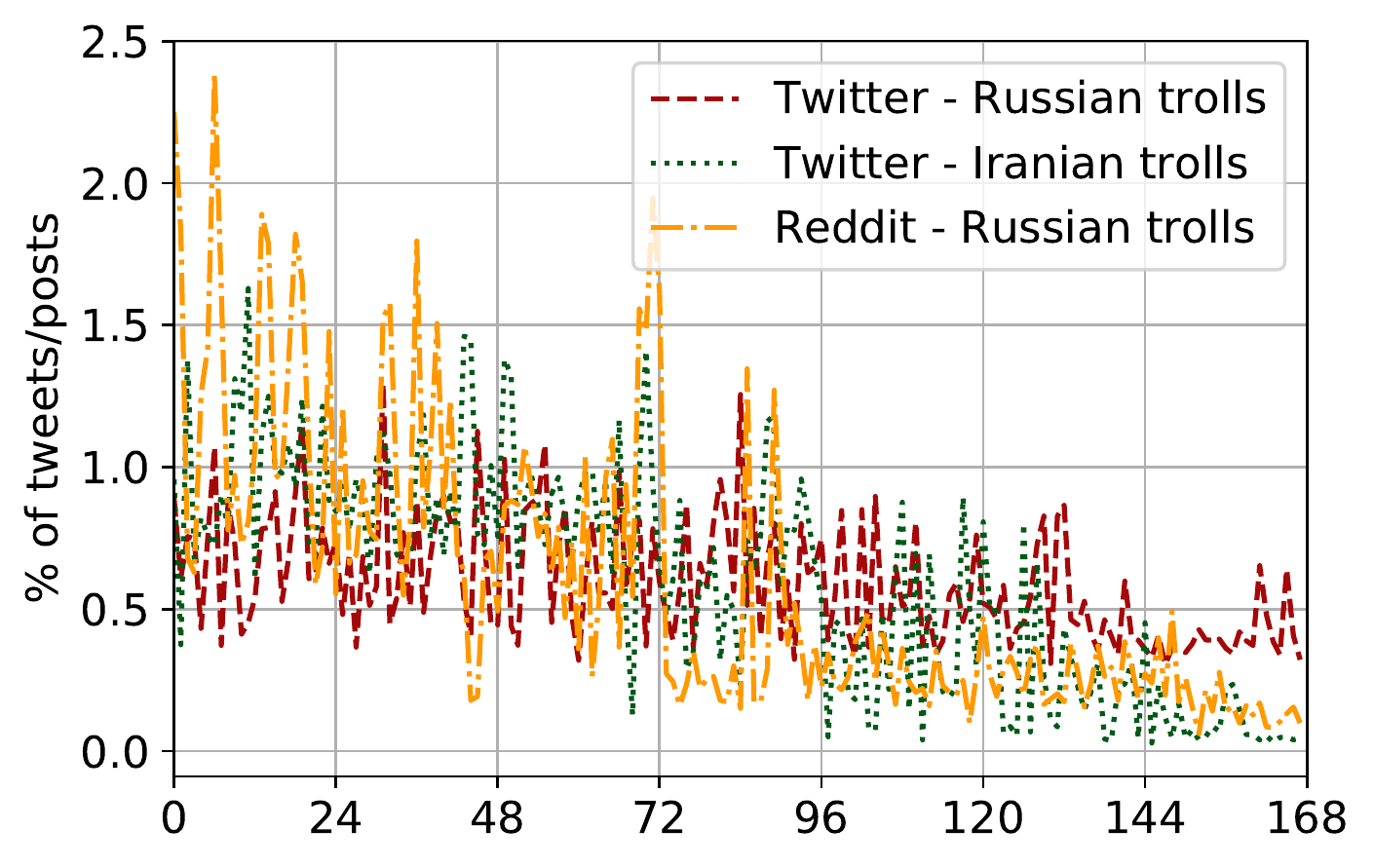}\label{subfig:counts_per_hour_week}}
  \caption{Temporal characteristics of tweets from Russian and Iranian trolls.} %
\label{fig:temporal_analysis}
\end{figure}

\begin{figure}[t!]
\centering
\includegraphics[width=0.97\columnwidth]{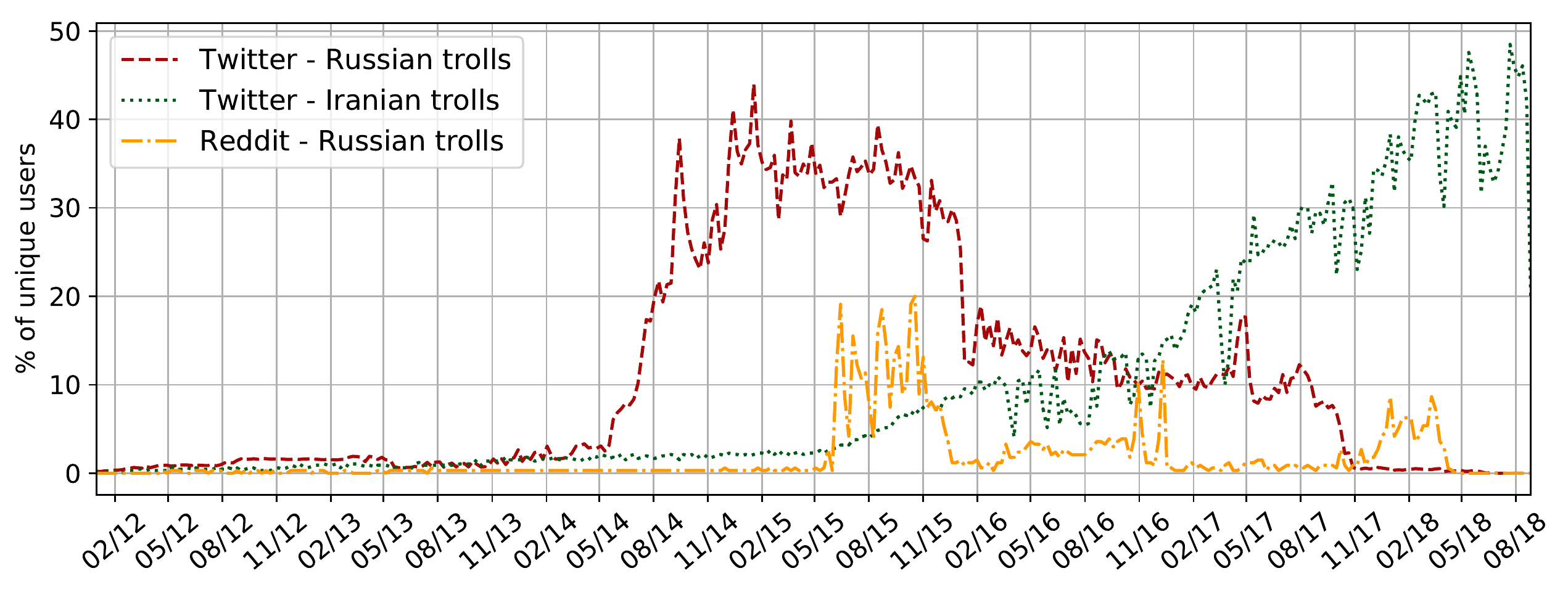}
\caption{Percentage of unique trolls that were active per week.}
\label{fig:unique_users_per_week}
\end{figure}

We next explore aggregate troll activity over time, looking for behavioral patterns.
Fig.~\ref{subfig:counts_per_day} plots the (normalized) volume of tweets/posts shared per week in our dataset.
We observe that both Russian and Iranian trolls on Twitter became active during the Ukrainian conflict.
Although lower in overall volume, there an increasing trend starts around August 2016 and continues through summer of 2017.

We also see three major spikes in activity by Russian trolls on Reddit. 
The first is during the latter half of 2015, approximately around the time that Donald Trump announced his candidacy for President.
Next, we see solid activity through the middle of 2016, trailing off shortly before the election.
Finally, we see another burst of activity in late 2017 through early 2018, at which point the trolls were detected and had their accounts locked by Reddit.

Next, we examine the hour of day and week that the trolls post.
Fig.~\ref{subfig:counts_per_hour_day} shows that Russian trolls on Twitter are active throughout the day, while on Reddit they are particularly active during the first hours of the day.
Similarly, Iranian trolls on Twitter tend to be active from early morning until 13:00 UTC.
In Fig.~\ref{subfig:counts_per_hour_week}, we report temporal characteristics based on hour of the week,
finding that Russian trolls on Twitter follow a diurnal pattern with slightly less activity during Sunday.
In contrast, Russian trolls on Reddit and Iranian trolls on Twitter are particularly active during the first days of the week, while their activity decreases during the weekend.
For Iranians this is likely due to the Iranian work week being from Sunday to Wednesday with a half day on Thursday.

\begin{figure}[t]
\centering
\includegraphics[width=0.97\columnwidth]{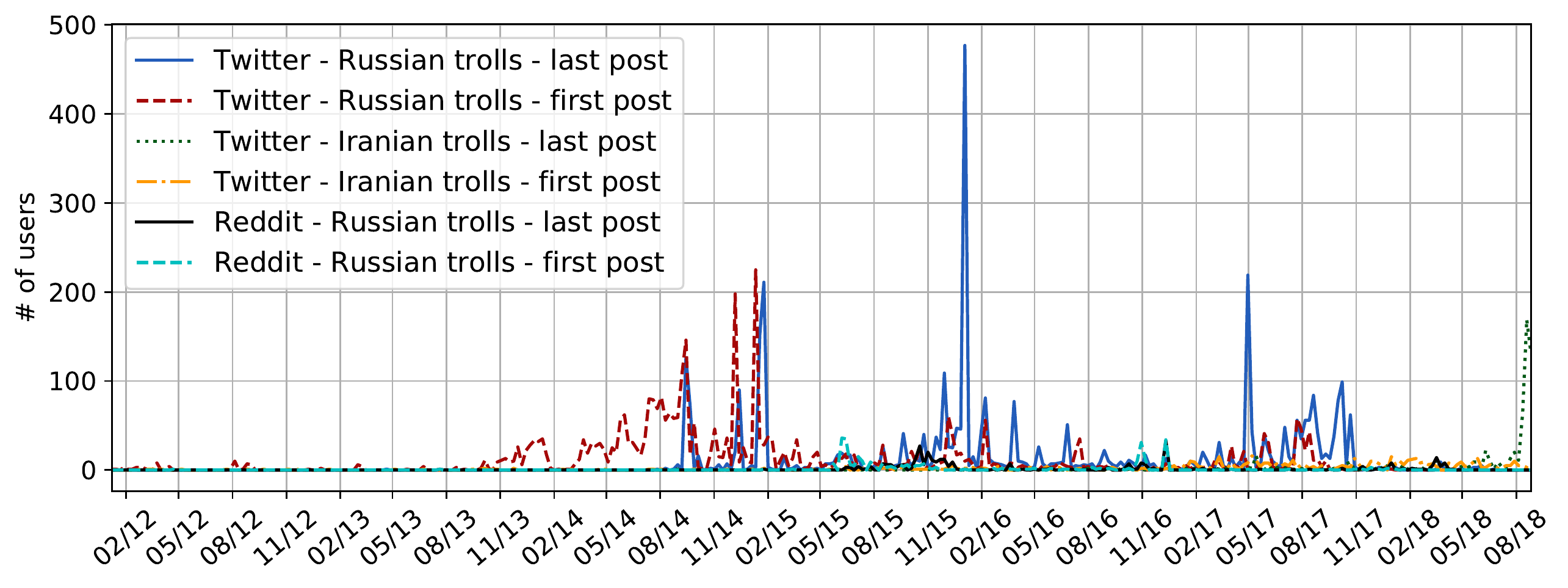}
\caption{Number of trolls that posted their first/last tweet/post for each week in our dataset.}
\label{fig:first_last_tweet}
\end{figure}

But are \emph{all} trolls in our dataset active throughout the span of our datasets?
To answer this question, we plot the percentage of unique troll accounts that are active per week in Fig.~\ref{fig:unique_users_per_week} from which we draw the following observations.
First, the Russian troll campaign on Twitter targeting Ukraine was much more diverse in terms of accounts when compared to later campaigns.
There are several possible explanations for this.
One explanation is that trolls learned from their Ukrainian campaign and became more efficient in later campaigns, perhaps relying on large networks of bots in their earlier campaigns which were later abandoned in favor of more focused campaigns like project Lakhta~\cite{lakhtaindictment}.
Another explanation could be that attacks on the US election might have required ``better trained'' trolls, perhaps those that could speak English more convincingly.
The Iranians, on the other hand, seem to be slowly building their troll army over time.
There is a steadily increasing number of active trolls posting per week over time.
We speculate that this is due to their troll program coming online in a slow-but-steady manner, perhaps due to more effective training.
Finally, on Reddit we see most Russian trolls posted irregularly, possibly performing other operations on the platform like manipulating votes on other posts.

Next, we investigate the point in time when each troll in our dataset made his first and last tweet.
Fig.~\ref{fig:first_last_tweet} shows the number of users that made their first/last post for each week in our dataset, which highlights when trolls became active as well as when they ``retired.''
We see that Russian trolls on Twitter made their first posts during early 2014, almost certainly in response to the Ukrainian conflict.
When looking at the last tweets of Russian trolls on Twitter we see that a substantial portion of the trolls ``retired'' by the end of 2015.
In all likelihood this is because the Ukrainian conflict was over and Russia turned their information warfare arsenal to other targets (e.g., the USA, this is also aligned with the increase in the use of English language, see Section~\ref{sec:language}).
When looking at Russian trolls on Reddit, we do not see a substantial spike in first posts close to the time that the majority of the accounts were created (see Fig.~\ref{fig:counts_created}).
This indicates that the newly created Russian trolls on Reddit became active gradually (in terms of posting behavior).

\begin{figure}[t!]
\centering
\includegraphics[width=0.97\columnwidth]{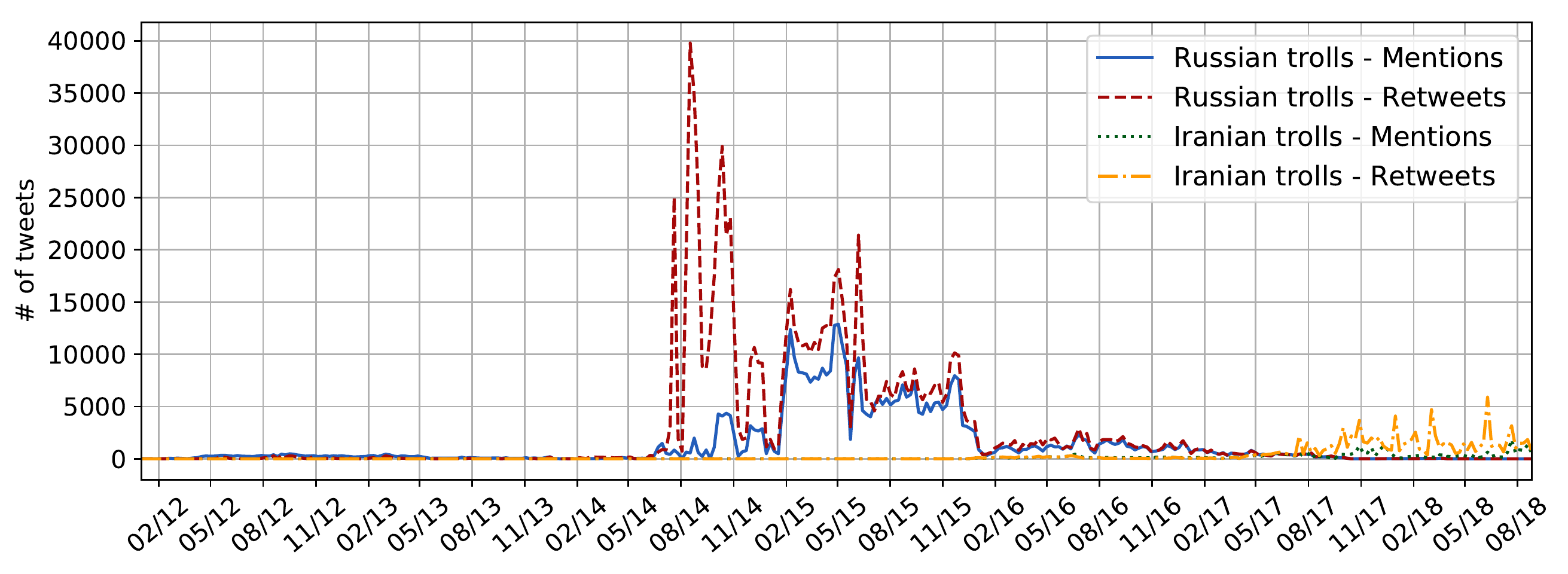}
\caption{Number of tweets that contain mentions among Russian trolls and among Iranian trolls on Twitter.}
\label{fig:mentions_among}
\end{figure}

Finally, we assess whether Russian and Iranian trolls mention or retweet each other, and how this behavior occurs over time.
Fig.~\ref{fig:mentions_among} shows the number of tweets that were mentioning/retweeting other trolls' tweets over the course of our datasets.
Russian trolls were particularly fond of this strategy during 2014 and 2015, while Iranian trolls started using this strategy after August, 2017.
This again highlights how the strategies employed by trolls adapts and evolves to new campaigns.

\subsection{Languages and Clients} \label{sec:language}

In this section, we study the languages that Russian and Iranian Twitter trolls posted in, as well as their Twitter clients they used to make tweets (this information is not available for Reddit).

\begin{figure}[t!]
\center
\subfigure[]{\includegraphics[width=0.485\columnwidth]{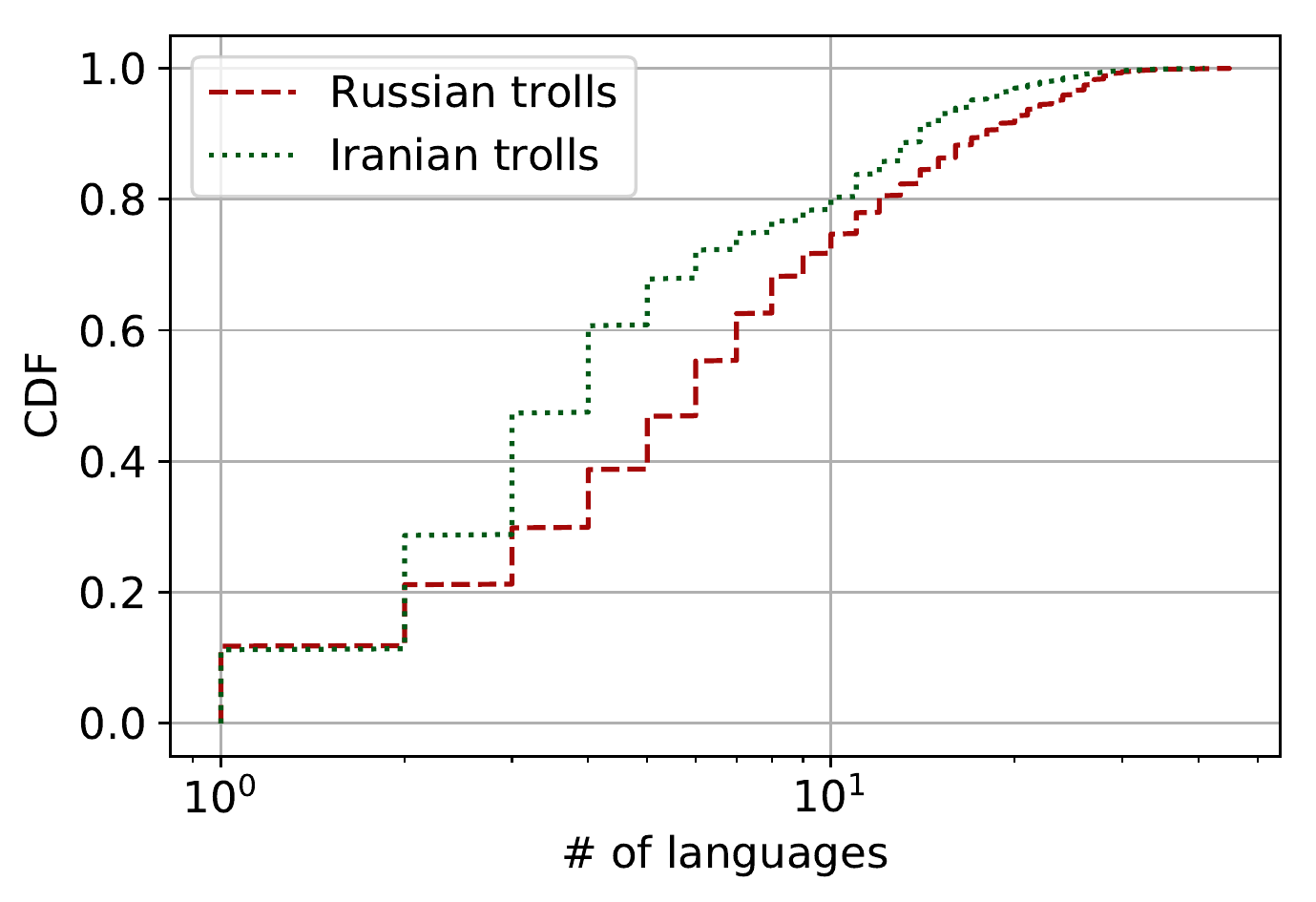}\label{subfig:cdf_languages_user}}
\subfigure[]{\includegraphics[width=0.485\columnwidth]{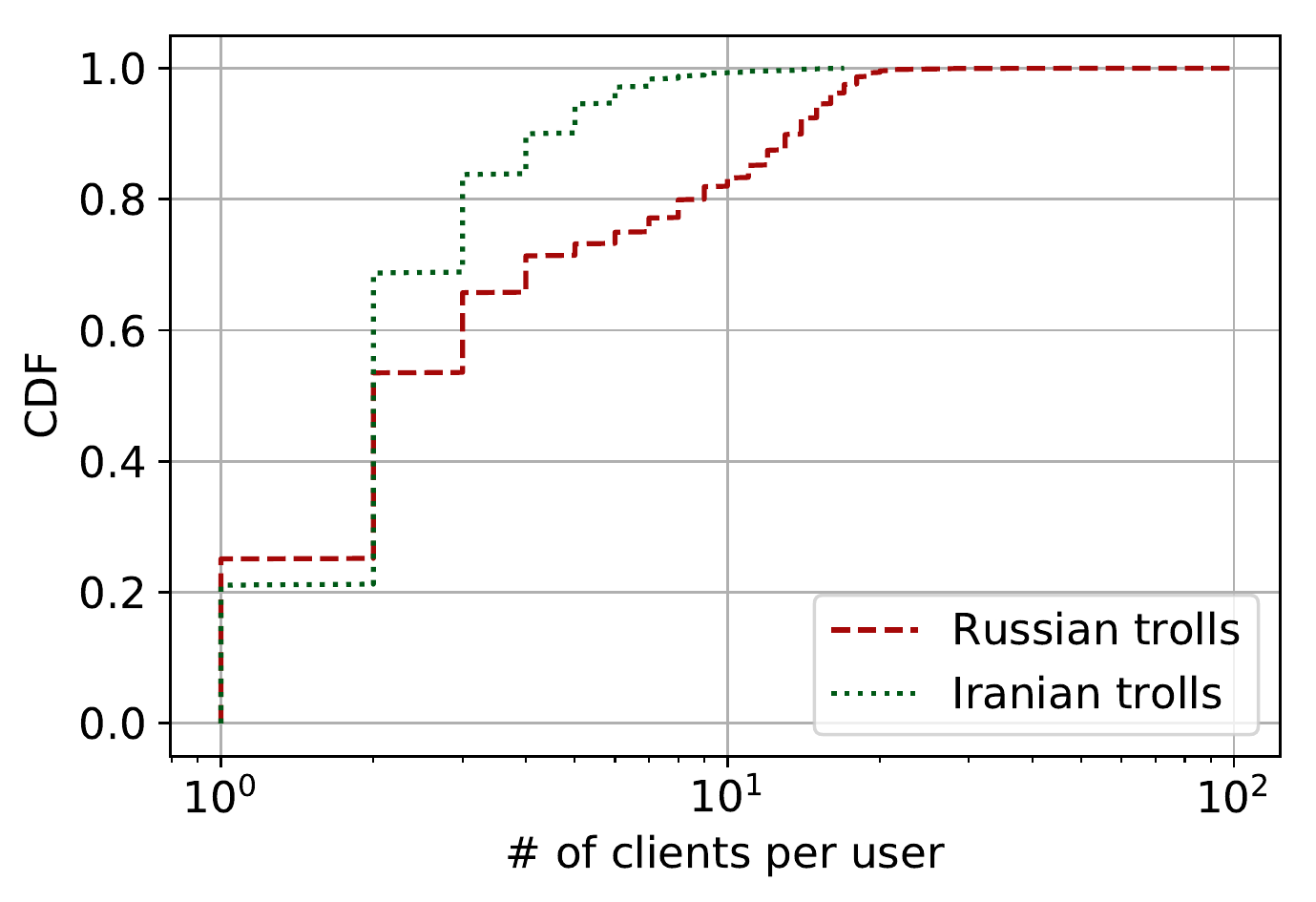}\label{subfig:cdf_sources_user}}
  \caption{CDF of number of (a) languages used (b) clients used for Russian and Iranian trolls on Twitter. }
\label{fig:cdf_lang_sources}
\end{figure}

\descr{Languages.}
First we study the languages used by trolls as it provides an indication of their targets.
The language information is included in the datasets released by Twitter.
Fig.~\ref{subfig:cdf_languages_user} plots the CDF of the number of languages used by troll accounts.
We find that 80\% and 75\% of the Russian and Iranian trolls, respectively, use more than 2 languages.
Next, we note that in general, Iranian trolls tend to use fewer languages than Russian trolls.
The most popular language for Russian trolls is Russian (53\% of all tweets), followed by English (36\%), Deutsch (1\%), and Ukrainian (0.9\%).
For Iranian trolls we find that French is the most popular language (28\% of tweets), followed by English (24\%), Arabic (13\%), and Turkish (8\%).

\begin{figure}[t!]
\center
\subfigure[Russians]{\includegraphics[width=0.97\columnwidth]{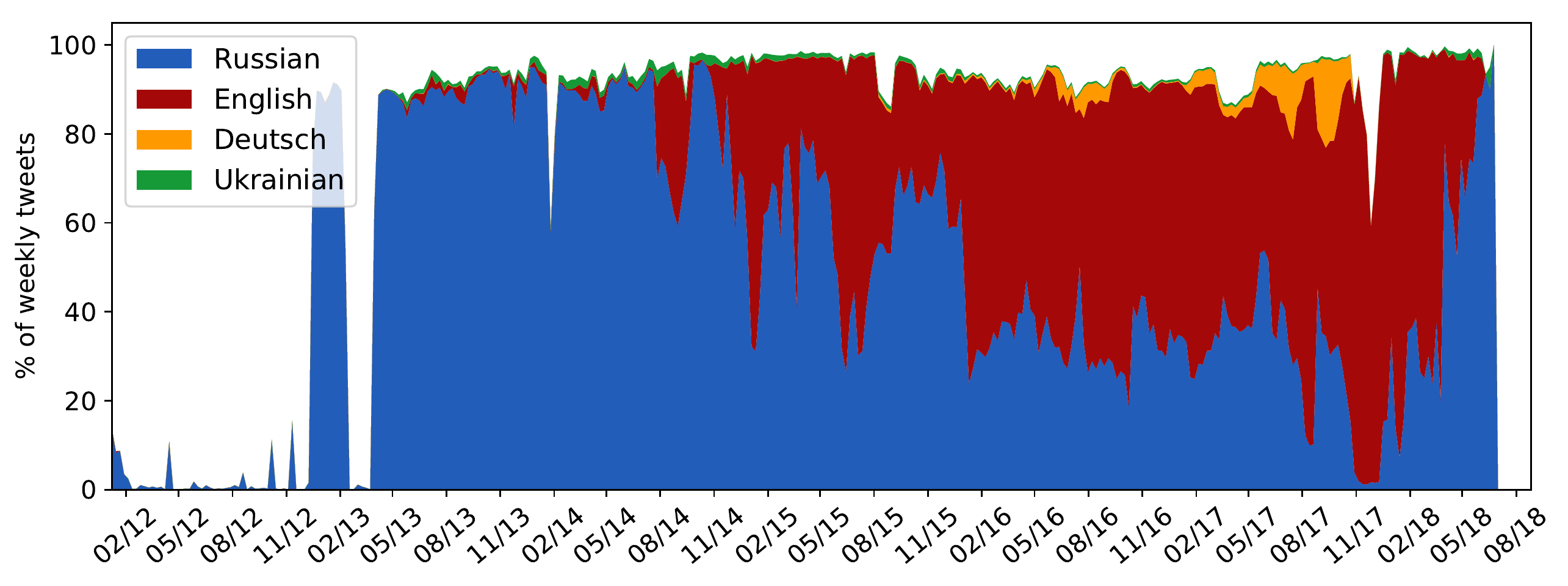}\label{subfig:russians_norm_week}}
\subfigure[Iranians]{\includegraphics[width=0.97\columnwidth]{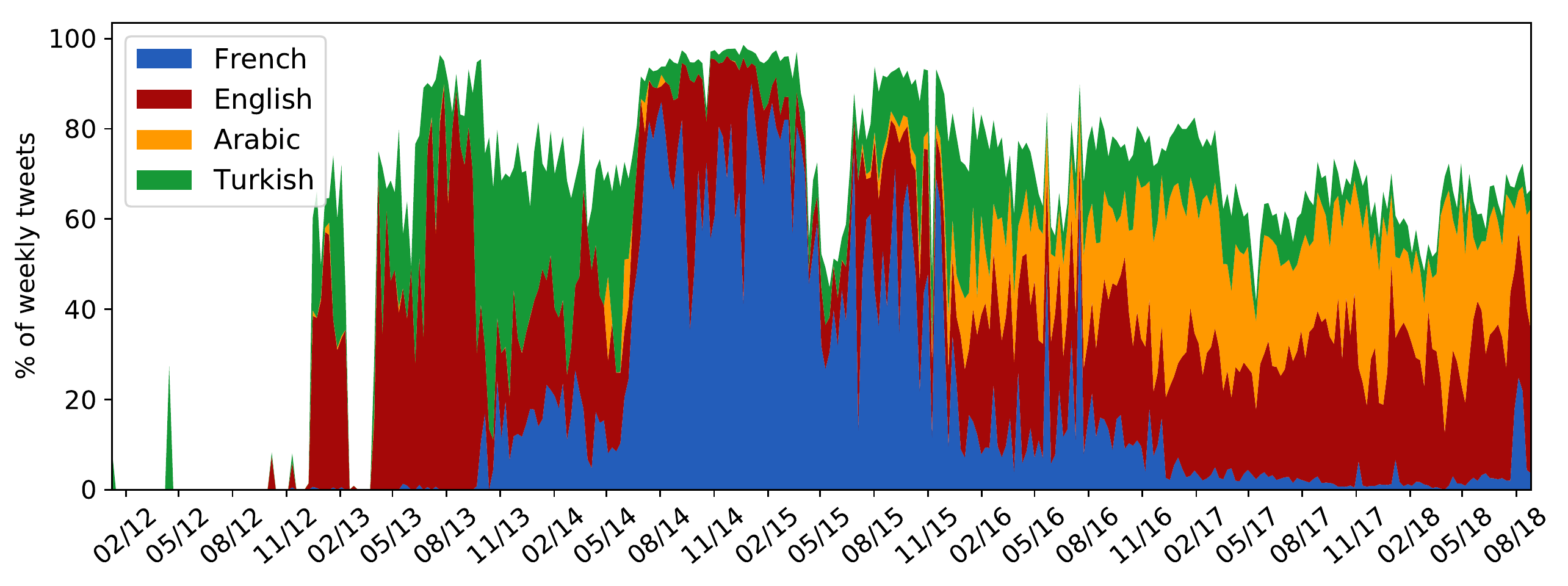}\label{subfig:iranians_norm_week}}
\subfigure[Russians]{\includegraphics[width=0.97\columnwidth]{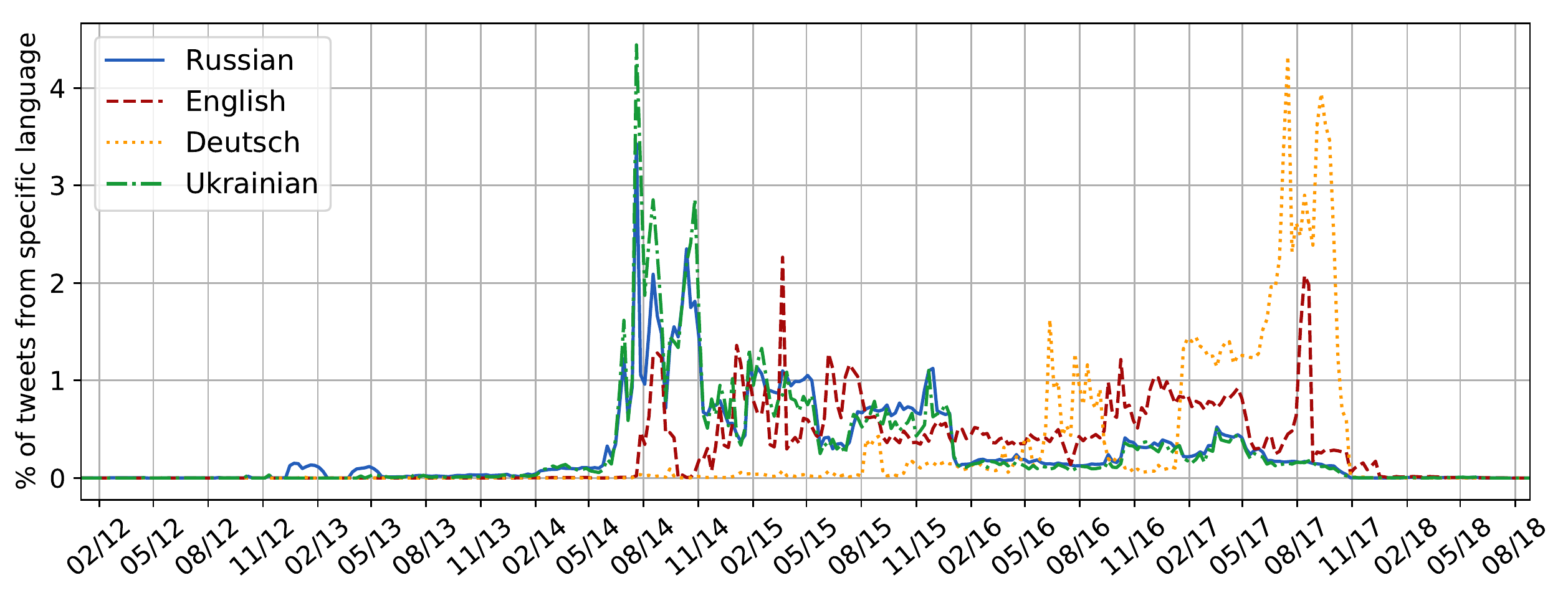}\label{subfig:russians_norm_language}}
\subfigure[Iranians]{\includegraphics[width=0.97\columnwidth]{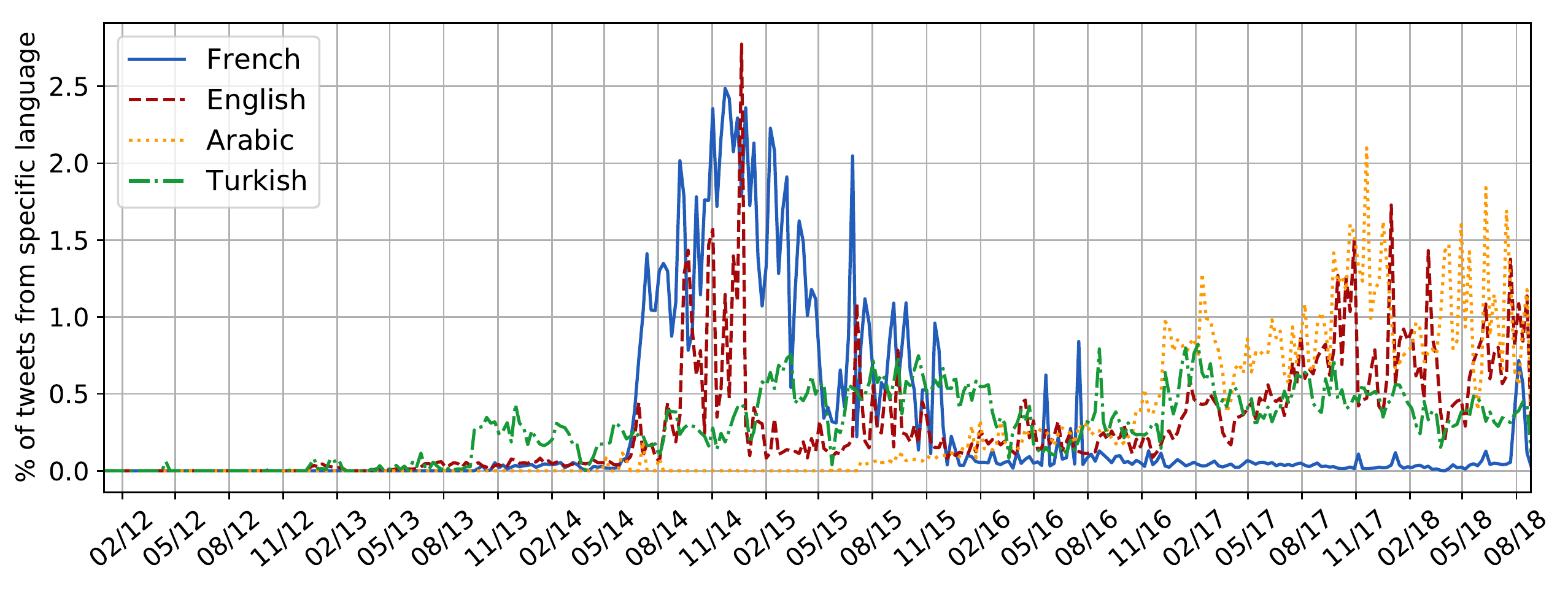}\label{subfig:iranians_norm_language}}
  \caption{Use of the four most popular languages by Russian and Iranian trolls over time on Twitter.
  (a) and (b) show the percentage of weekly tweets in each language.
  (c) and (d) show the percentage of total tweets per language that occurred in a given week.
  }
\label{fig:languages_over_time}
\end{figure}

Fig.~\ref{fig:languages_over_time} plots the use of different languages over time.
Fig.~\ref{subfig:russians_norm_week} and Fig.~\ref{subfig:iranians_norm_week} plot the percentage of tweets that were in a given language on a given week for Russian and Iranian trolls, respectively, in a stacked fashion, which lets us see how the usage of different languages changed over time relative to each other.
Fig.~\ref{subfig:russians_norm_language} and Fig.~\ref{subfig:iranians_norm_language} plot the language use from a different perspective: normalized to the overall number of tweets in a given language.
This view gives us a better idea of how the use of each particular language changed over time.
From the plots we make the following observations.
First, there is a clear shift in targets based on the campaign.
For example, Fig.~\ref{subfig:russians_norm_week} shows that the overwhelming majority of early tweets by Russian trolls were in Russian, with English only reaching the volume of Russian language tweets in 2016.
This coincides with the ``retirement'' of several Russian trolls on Twitter (see Fig~\ref{fig:first_last_tweet}).
Next, we see evidence of other campaigns, for example German language tweets begin showing up in early to mid 2016, and reach their highest volume in the latter half of 2017, in close proximity with the 2017 German Federal elections.
Additionally, we note that Russian language tweets have a huge drop off in activity the last two months of 2017.

For the Iranians, we see more obvious evidence of multiple campaigns.
For example, although Turkish and English are present for most of the timeline, French quickly becomes a commonly used language in the latter half of 2013, becoming the dominant language used from around May 2014 until the end of 2015.
This is likely due to political events that happened during this time period.
E.g., in November, 2013 France blocked a stopgap deal related to Iran's uranium enrichment program~\cite{guardian_iran}, leading to some fiery rhetoric from Iran's government (and apparently the launch of a troll campaign targeting French speakers).
As tweets in French fall off, we also observe a dramatic increase in the use of Arabic in early 2016.
This coincides with an attack on the Saudi embassy in Tehran~\cite{nytimes_iran_embassy}, the primary reason the two countries ended diplomatic relations.

When looking at the language usage normalized by the total number of tweets in that language, we can get a more focused perspective.
In particular, from Fig.~\ref{subfig:russians_norm_language} it becomes strikingly clear that the initial burst of Russian troll activity was targeted at Ukraine, with the majority of Ukrainian language tweets coinciding directly with the Crimean conflict~\cite{crimea_timeline}.
From Fig.~\ref{subfig:iranians_norm_language} we observe that English language tweets from Iranian trolls, while consistently present over time, have a relative peak corresponding with French language tweets, likely indicating an attempt to influence non-French speakers with respect to the campaign against French speakers.

\begin{figure}[t!]
\center
\subfigure[Russians]{\includegraphics[width=0.97\columnwidth]{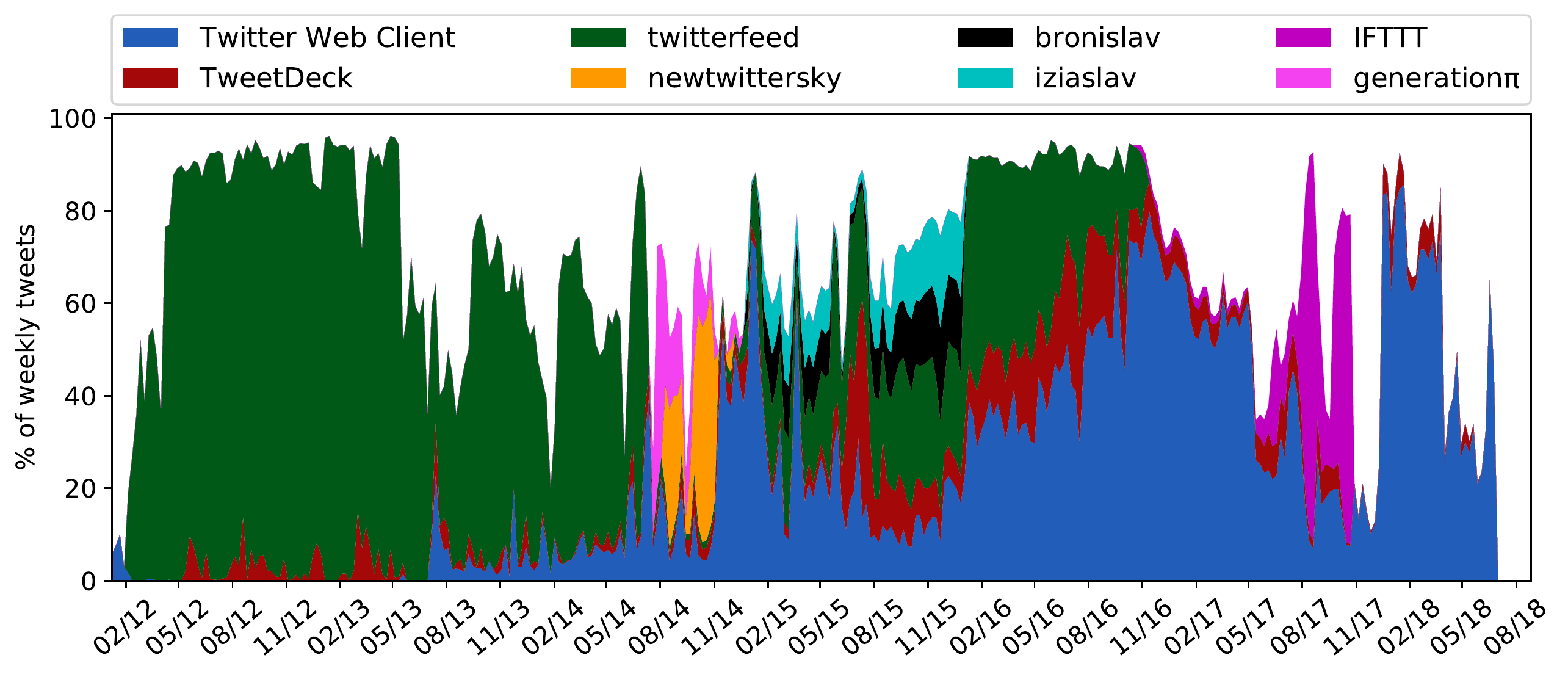}\label{subfig:russians_norm_week_clients}}
\subfigure[Iranians]{\includegraphics[width=0.97\columnwidth]{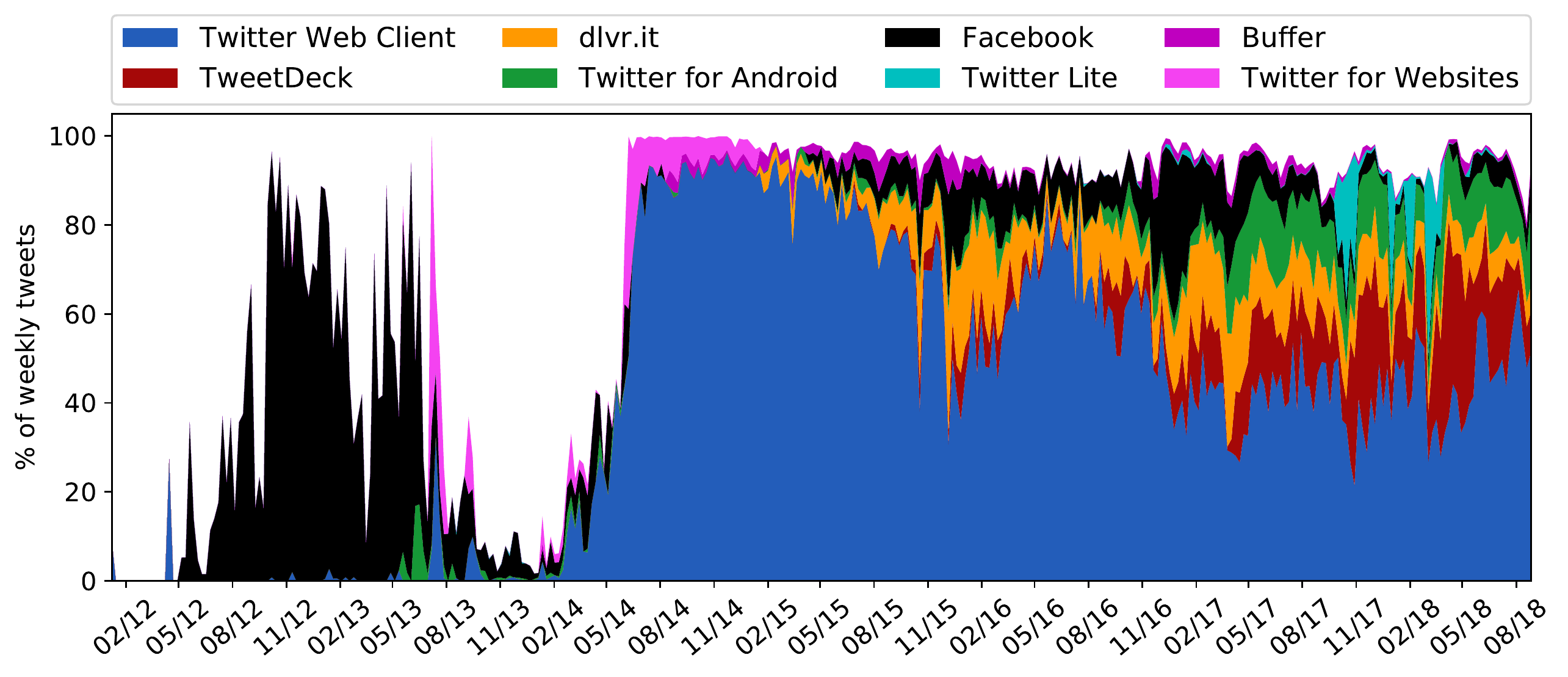}\label{subfig:iranians_norm_week_clients}}
  \caption{Use of the eight most popular clients by Russian and Iranian trolls over time on Twitter.
  }
\label{fig:clients_over_time}
\end{figure}

\begin{figure*}[t!]
\centering
\includegraphics[width=0.7\textwidth]{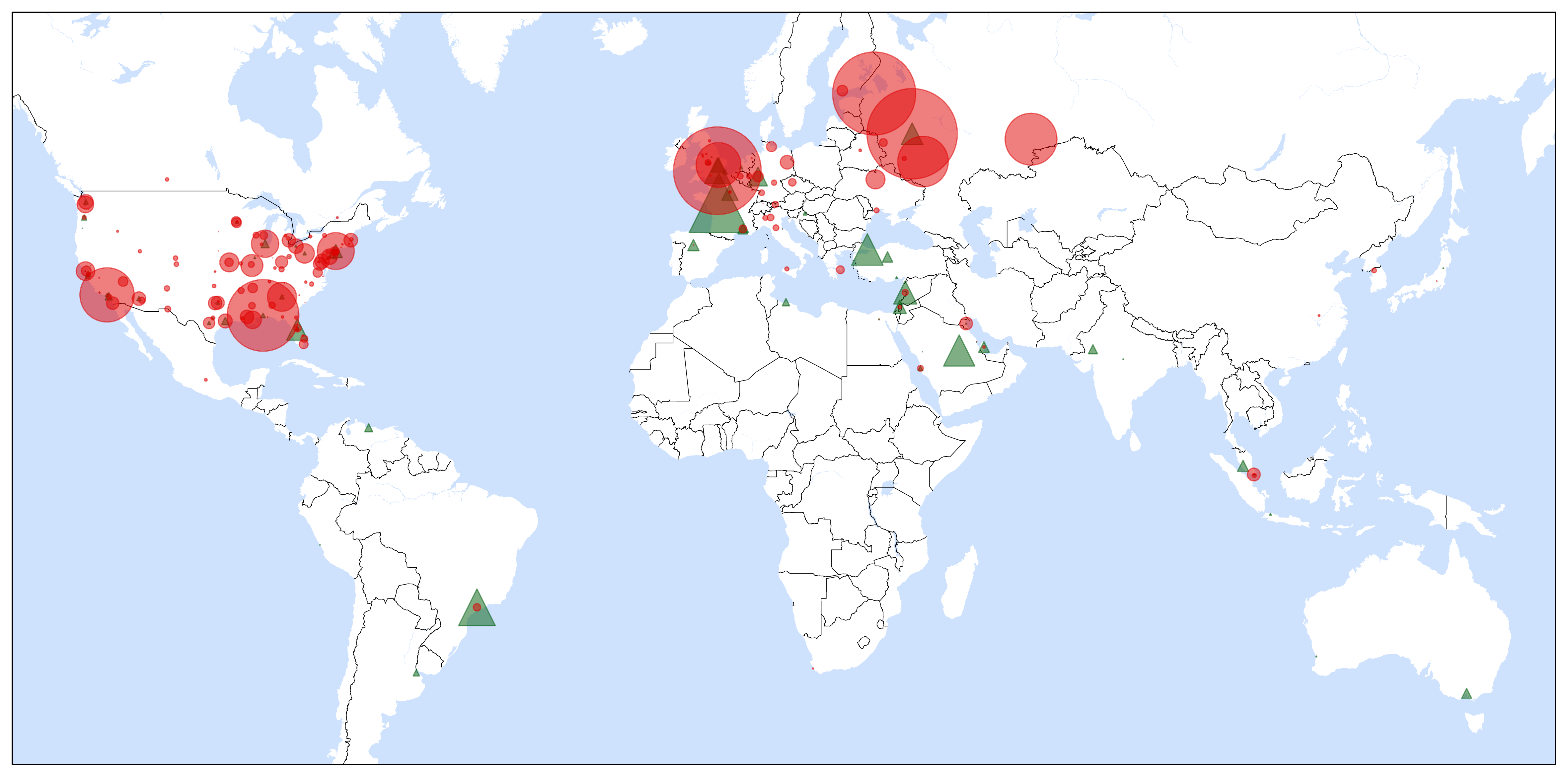}
   \caption{Distribution of reported locations for tweets by Russian trolls (100\%) (red circles) and Iranian trolls (green triangles).}
\label{fig:locations_map_agg}
\end{figure*}
\descr{Client usage.}
Finally, we analyze the clients used to post tweets.
When looking at the most popular clients, we find that Russian and Iranian trolls use the main Twitter Web Client (28.5\% for Russian trolls, and 62.2\% for Iranian trolls).
This is in contrast with what normal users use: using a random set of Twitter users, we find that mobile clients make up a large chunk of tweets (48\%), followed by the TweetDeck dashboard (32\%).
We next look at how many different clients trolls use throughout our dataset:
in Fig.~\ref{subfig:cdf_sources_user}, we plot the CDF of the number of clients used per user.
25\% and 21\% of the Russian and Iranian trolls, respectively, use only one client, while in general Russian trolls tend to use more clients than Iranians.

Fig.~\ref{fig:clients_over_time} plots the usage of clients over time in terms of weekly tweets by Russian and Iranian trolls.
We observe that the Russians (Fig.~\ref{subfig:russians_norm_week_clients}) started off with almost exclusive use of the ``twitterfeed'' client.
Usage of this client drops off when it was shutdown in October, 2016.
During the Ukrainian crisis, however, we see several new clients come into the mix.
Iranians (Fig.~\ref{subfig:iranians_norm_week_clients}) started off almost exclusively using the ``facebook'' Twitter client.
To the best of our knowledge, this is a client that automatically Tweets any posts you make on Facebook, indicating that Iranians likely started with a campaign on Facebook.
At the beginning of 2014, we see a shift to using the Twitter Web Client, which only begins to decrease towards the end of 2015.
Of particular note in Fig.~\ref{subfig:iranians_norm_week_clients} is the appearance of ``dlvr.it,'' an automated social media manager, in the beginning of 2015.
This corresponds with the creation of IUVM~\cite{iuvm_about_page}, which is a fabricated ecosystem of (fake) news outlets and social media accounts created by the Iranians, and might indicate that Iranian trolls stepped up their game around that time, starting using services that allowed them for better account orchestration to run their campaigns more effectively.

\subsection{Geographical Analysis}

We then study users' location, relying on the self-reported location field in their profiles, since only very few tweets have actual GPS coordinates.
Note that this field is not required, and users are also able to change it whenever they like,
so we look at locations for each tweet.
Note that 16.8\% and 20.9\% of the tweets from Russian and Iranians trolls, respectively, do not include a self-reported location.
To infer the geographical location from the self-reported text, we use pigeo~\cite{rahimi2016pigeo}, which provides geographical information (e.g., latitude, longitude, country, etc.) given the text that corresponds to a location.
Specifically, we extract 626 self-reported locations for the Russian trolls and 201 locations for the Iranian trolls.
Then, we use pigeo to systematically obtain a geographical location (and its associated coordinates) for each text that corresponds to a location.
Fig.~\ref{fig:locations_map_agg} shows the locations inferred for Russian trolls (red circles) and Iranian trolls (green triangles).
The size of the shapes on the map indicates the number of tweets that appear on each location.
We observe that most of the tweets from Russian trolls come from locations within Russia (34\%), the USA (29\%), and some from European countries, like United Kingdom (16\%), Germany (0.8\%), and Ukraine (0.6\%).
This suggests that Russian trolls may be pretending to be from certain countries, e.g., USA or United Kingdom, aiming to pose as locals and effectively manipulate opinions.
A similar pattern exists with Iranian trolls, which were particularly active in France (26\%), Brazil (9\%), the USA (8\%), Turkey (7\%), and Saudi Arabia (7\%).
It is also worth noting that Iranians trolls, unlike Russian trolls, did not report locations from their country, indicating that these trolls were primarily used for campaigns targeting foreign countries.
Finally, we note that the location-based findings are in-line with the findings on the languages analysis (see Section~\ref{sec:language}), further evidencing that both Russian and Iranian trolls were specifically targeting different countries over time.

\subsection{Content Analysis}

\begin{table}[]
\centering
\resizebox{0.75\columnwidth}{!}{%
\begin{tabular}{@{}lrlr@{}}
\toprule
\multicolumn{2}{c}{\textbf{Russian trolls on Twitter}} & \multicolumn{2}{c}{\textbf{Iranian trolls on Twitter}} \\ \midrule
\textbf{Word} & \multicolumn{1}{l}{\textbf{\begin{tabular}[c]{@{}l@{}}Cosine\\ Similarity\end{tabular}}} & \textbf{Word} & \multicolumn{1}{l}{\textbf{\begin{tabular}[c]{@{}l@{}}Cosine \\ Similarity\end{tabular}}} \\ \midrule
trumparmi & \multicolumn{1}{r|}{0.68} & impeachtrump & 0.81 \\
trumptrain & \multicolumn{1}{r|}{0.67} & stoptrump & 0.80 \\
votetrump & \multicolumn{1}{r|}{0.65} & fucktrump & 0.79 \\
makeamericagreatagain & \multicolumn{1}{r|}{0.65} & trumpisamoron & 0.79 \\
draintheswamp & \multicolumn{1}{r|}{0.62} & dumptrump & 0.79 \\
trumppenc & \multicolumn{1}{r|}{0.61} & ivankatrump & 0.77 \\
@realdonaldtrump & \multicolumn{1}{r|}{0.59} & theresist & 0.76 \\
wakeupamerica & \multicolumn{1}{r|}{0.58} & trumpresign & 0.76 \\
thursdaythought & \multicolumn{1}{r|}{0.57} & notmypresid & 0.76 \\
realdonaldtrump & \multicolumn{1}{r|}{0.57} & worstpresidentev & 0.75 \\
presidenttrump & \multicolumn{1}{r|}{0.57} & antitrump & 0.74 \\ \bottomrule
\end{tabular}%
}
\caption{Top 10 similar words to ``maga'' and their respective cosine similarities (obtained from the word2vec models).}
\label{tbl:similar_to_maga}
\end{table}

\begin{table}[]
\centering
\setlength{\tabcolsep}{0.2em} %
\hspace*{-0.2cm}
\resizebox{1.03\columnwidth}{!}{
\begin{tabular}{lrlrlrlr}
\hline
\multicolumn{4}{c}{\textbf{Russian trolls on Twitter}} & \multicolumn{4}{c}{\textbf{Iranian trolls on Twitter}} \\
\textbf{Hashtag} & \textbf{(\%)} & \textbf{Hashtag} & \multicolumn{1}{l}{\textbf{(\%)}} & \textbf{Hashtag} & \textbf{(\%)} & \textbf{Hashtag} & \textbf{(\%)} \\ \hline
news & 9.5\% & USA & \multicolumn{1}{r|}{0.7\%} & Iran & 1.8\% & Palestine & 0.6\% \\
sports & 3.8\% & breaking & \multicolumn{1}{r|}{0.7\%} & Trump & 1.4\% & Syria & 0.5\% \\
politics & 3.0\% & TopNews & \multicolumn{1}{r|}{0.6\%} & Israel & 1.1\% & Saudi & 0.5\% \\
local & 2.1\% & BlackLivesMatter & \multicolumn{1}{r|}{0.6\%} & Yemen & 0.9\% & EEUU & 0.5\% \\
world & 1.1\% & true & \multicolumn{1}{r|}{0.5\%} & FreePalestine & 0.8\% & Gaza & 0.5\% \\
MAGA & 1.1\% & Texas & \multicolumn{1}{r|}{0.5\%} & QudsDay4Return & 0.8\% & SaudiArabia & 0.4\% \\
business & 1.0\% & NewYork & \multicolumn{1}{r|}{0.4\%} & US & 0.7\% & Iuvm & 0.4\% \\
Chicago & 0.9\% & Fukushima2015 & \multicolumn{1}{r|}{0.4\%} & realiran & 0.6\% & InternationalQudsDay2018 & 0.4\% \\
health & 0.8\% & quote & \multicolumn{1}{r|}{0.4\%} & ISIS & 0.6\% & Realiran & 0.4\% \\
love & 0.7\% & Foke & \multicolumn{1}{l|}{0.4\%} & DeleteIsrael & 0.6\% & News & 0.4\% \\ \hline
\end{tabular}
}
\caption{Top 20 (English) hashtags in tweets from Russian and Iranian trolls on Twitter.}
\label{tbl:top_hashtags_russians_iranians}
\end{table}

\begin{figure*}[t!]
\centering
\subfigure[]{\includegraphics[width=0.49\textwidth]{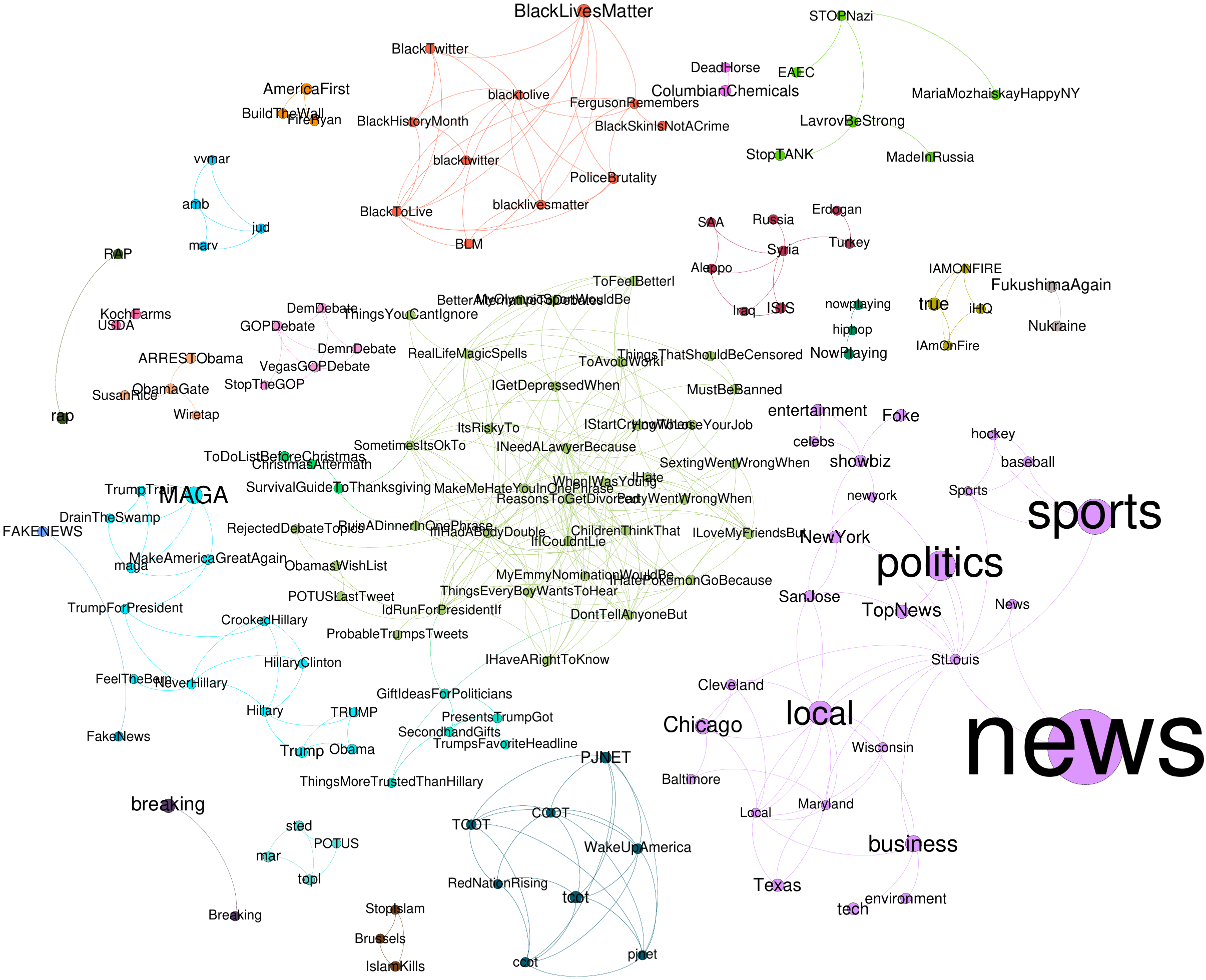}\label{fig:russians_hashtags_graph}}
\subfigure[]{\includegraphics[width=0.49\textwidth]{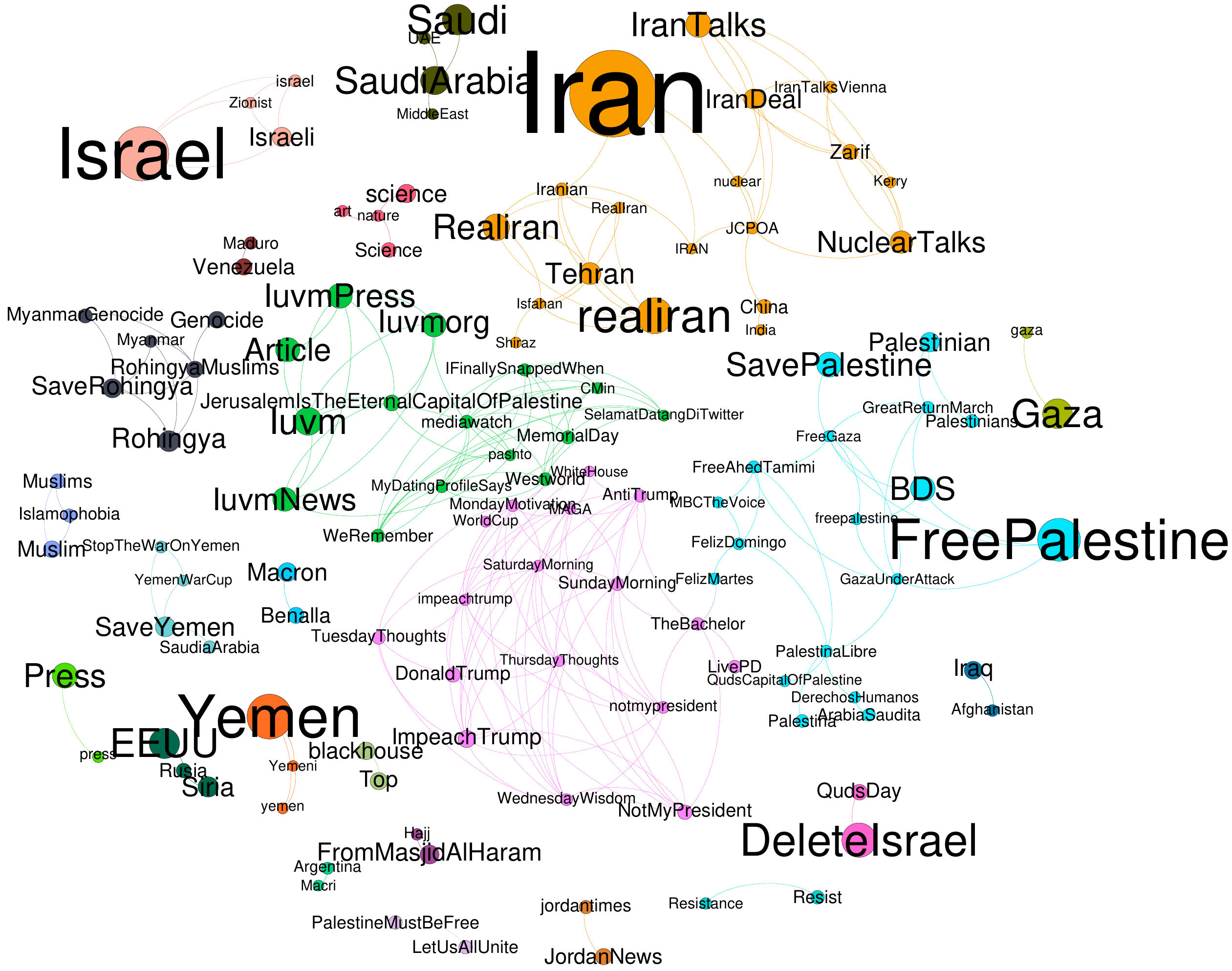} \label{fig:iranians_hashtags_graph}}
   \caption{Visualization of the top hashtags used by a) Russian trolls on Twitter (see~\cite{interactive_russians} for interactive version) and b) Iranian trolls on Twitter (see~\cite{interactive_iranians} for an interactive version).}
\label{fig:hashtags_graphs}
\end{figure*}

\descr{Word Embeddings.}
Recent indictments by the US Department of Justice have indicated that troll messaging was crafted, with certain phrases and terminology designated for use in certain contexts. 
To get a better handle on how this was expressed, we build two word2vec models on the corpus of tweets: one for the Russian trolls and one for the Iranian trolls.
To train the models, we first extract the tweets posted in English, according to the data provided by Twitter. 
Then, we remove stop words, perform stemming, tokenize the tweets, and keep only words that appear at least 500 and 100 times for the Russian and Iranian trolls, respectively.

Table~\ref{tbl:similar_to_maga} shows the top 10 most similar terms to ``maga'' for each model.
We see a marked difference between its usage by Russian and Iranian trolls.
Russian trolls are clearly pushing heavily in favor of Donald Trump, while it is the exact opposite with Iranians.

\descr{Hashtags.} Next, we aim to understand the use of hashtags with a focus on the ones written in English.
In Table~\ref{tbl:top_hashtags_russians_iranians}, we report the top 20 English hashtags for both Russian and Iranian trolls.
State-sponsored trolls appear to use hashtags to disseminate news (9.5\%) and politics (3.0\%) related content, but also use several that might be indicators of propaganda and/or controversial topics, e.g., \#BlackLivesMatter.
For instance, one notable example is:
``WATCH: Here is a typical \#BlackLivesMatter protester:  `I hope I kill all white babes!' \#BatonRouge $<$url$>$'' on July 17, 2016. Note that $<$url$>$ denotes a link.

Fig.~\ref{fig:hashtags_graphs} shows a visualization of hashtag usage built from the two word2vec models.
Here, we show hashgtags used in a similar context, by constructing a graph where nodes are words that correspond to hashtags from the word2vec models, and edges are weighted by the cosine distances (as produced by the word2vec models) between the hashtags.
After trimming out all edges between nodes with weight less than a threshold, based on methodology from~\cite{finkelstein2018quantitative}, we run the community detection heuristic presented in~\cite{blondel2008fast}, and mark each community with a different color.
Finally, the graph is layed out with the ForceAtlas2 algorithm~\cite{jacomy2014forceatlas2}, which takes into account the weight of the edges when laying out the nodes in 2-dimensional space.
Note that the size of the nodes is proportional to the number of times the hashtag appeared in each dataset.

We first observe that, in Fig.~\ref{fig:russians_hashtags_graph} there is a central mass of what we consider ``general audience'' hashtags (see green community on the center of the graph): hashtags related to a holiday or a specific trending topic (but non-political) hashtag.
In the bottom right portion of the plot we observe ``general news'' related categories; in particular American sports related hashtags (e.g., ``baseball'').
Next, we see a community of hashtags (light blue, towards the bottom left of the graph) clearly related to Trump's attacks on Hillary Clinton.

The Iranian trolls again show different behavior.
There is a community of hashtags related to nuclear talks (orange), a community related to Palestine (light blue), and a community that is clearly anti-Trump (pink).
The central green community exposes some of the ways they pushed the IUVM fake news network by using innocuous hashtags like ``\#MyDatingProfileSays'' as well as politically motivated ones like ``\#JerusalemIsTheEternalCapitalOfPalestine.''

\begin{figure*}[t!]
\center
\subfigure[Russian trolls - Before election]{\includegraphics[width=0.97\columnwidth]{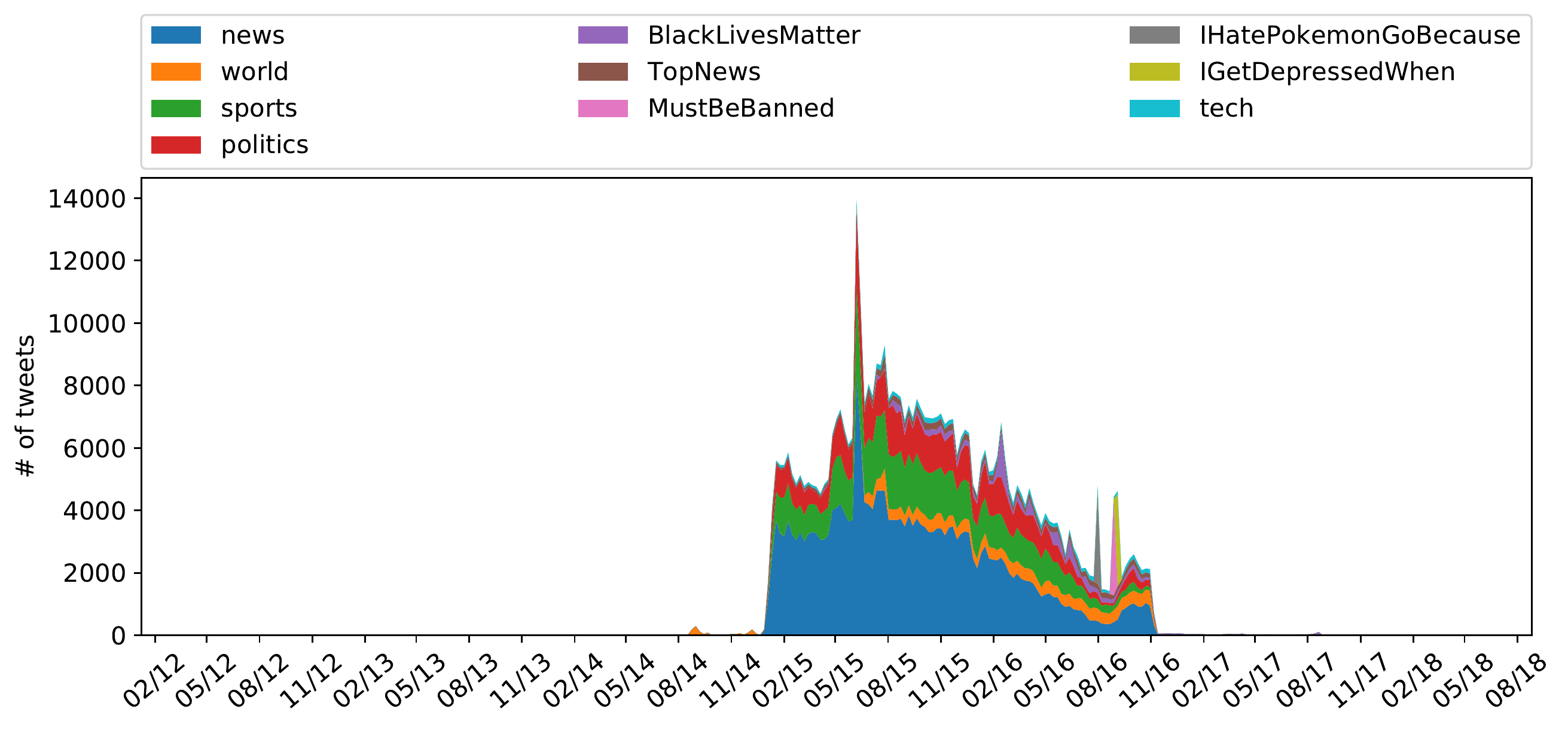}\label{subfig:russians_elections_before}}
\subfigure[Russian trolls - After election]{\includegraphics[width=0.97\columnwidth]{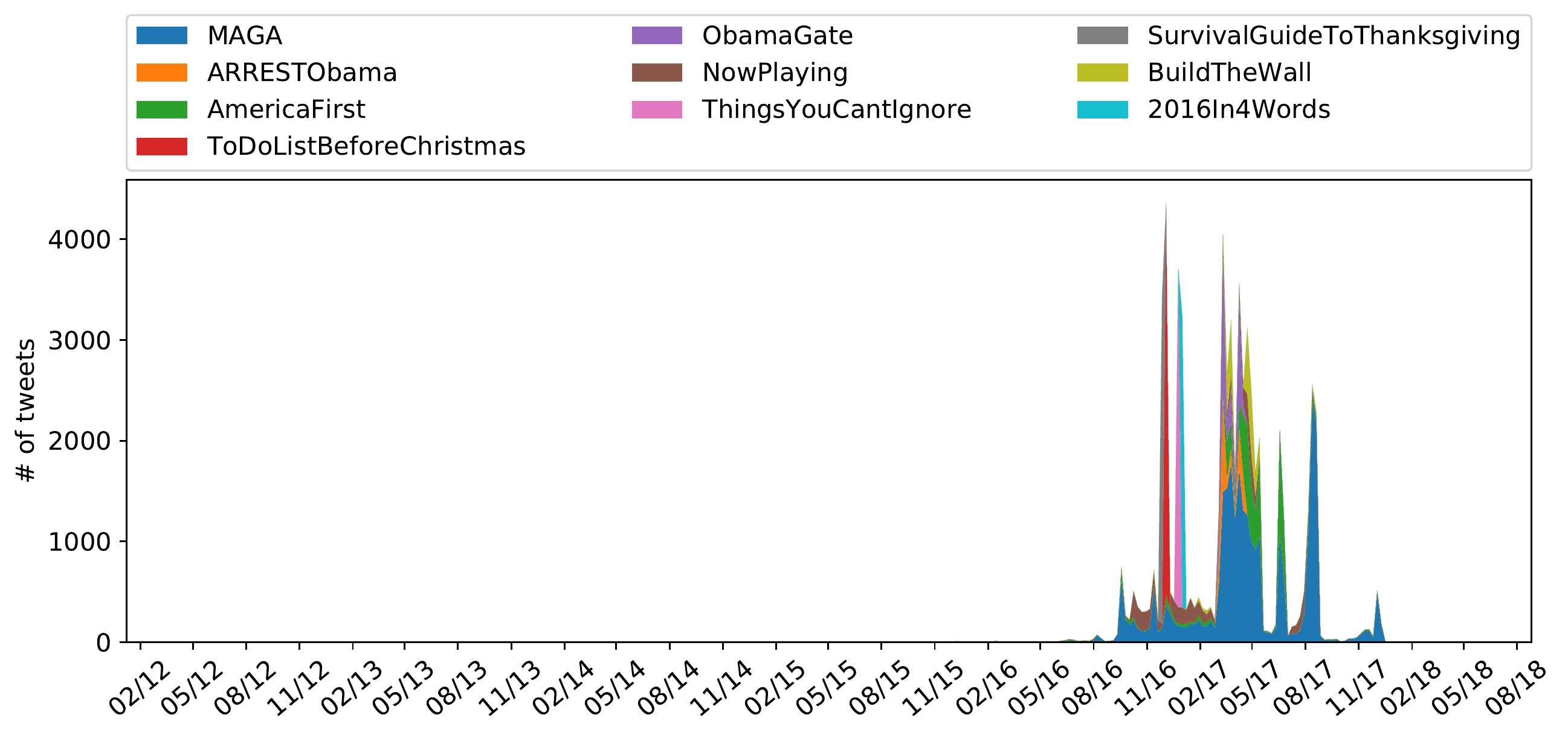}\label{subfig:russians_elections_after}}
\subfigure[Iranians trolls - Before election]{\includegraphics[width=0.97\columnwidth]{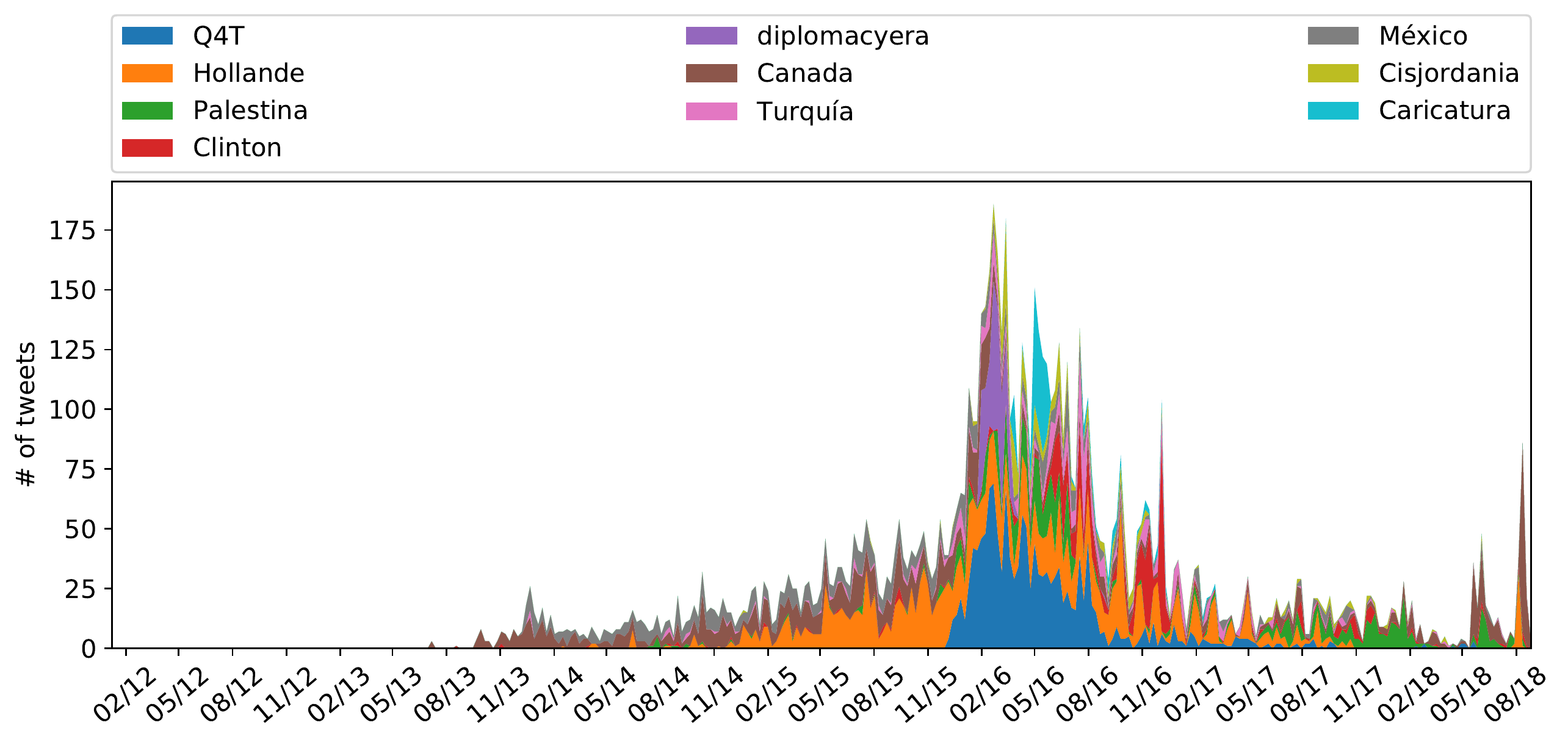}\label{subfig:iranians_elections_before}}
\subfigure[Iranians trolls - After election]{\includegraphics[width=0.97\columnwidth]{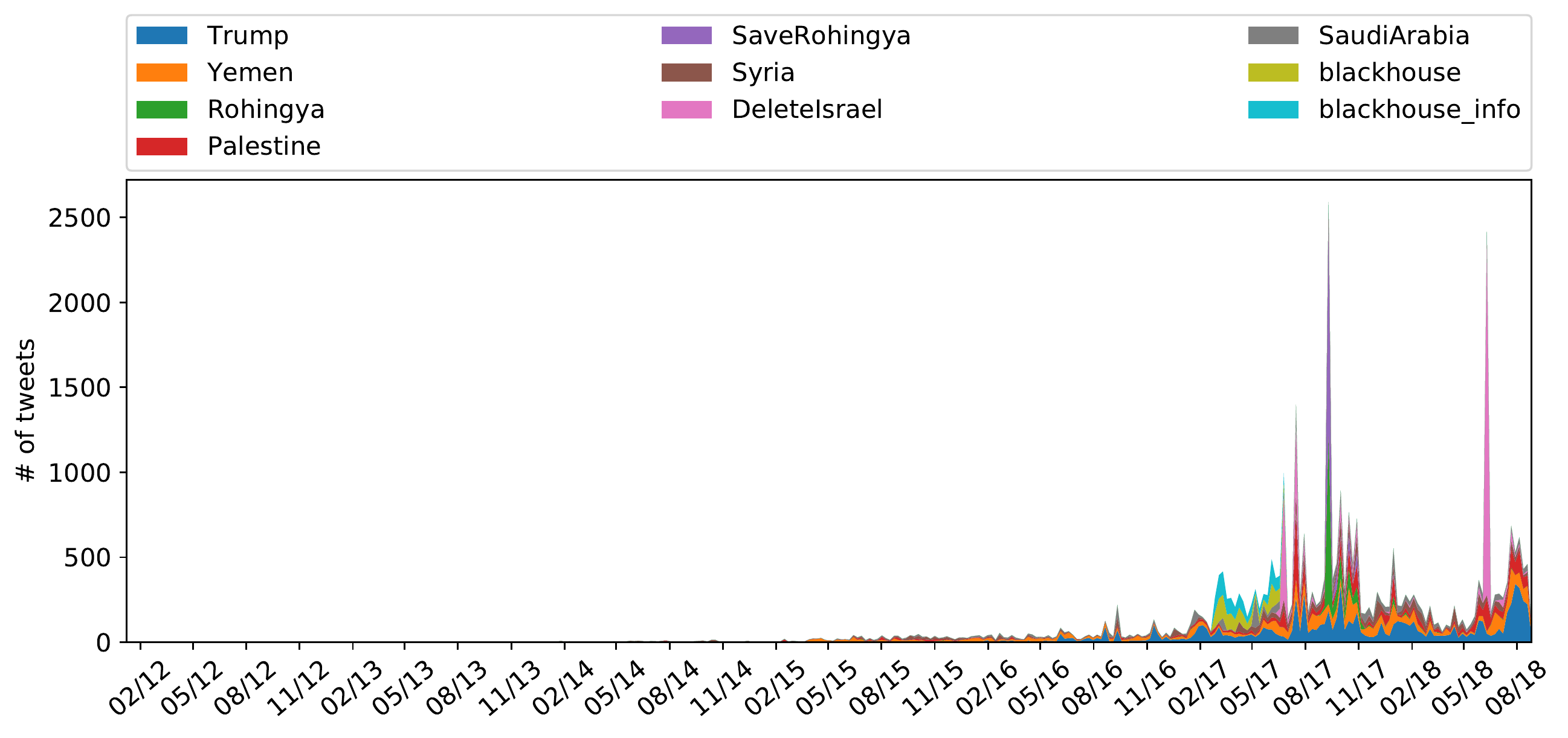}\label{subfig:iranians_elections_after}}
  \caption{ Top ten hashtags that appear a) c) substantially more times before the US elections rather than after the elections; and b) d) substantially more times after the elections rather than before. }
\label{fig:hashtags_before_after_election_russians}
\end{figure*}

We also study \emph{when} these hashtags are used by the trolls, finding that most of them are well distributed over time.
However we find some interesting exceptions.
We highlight a few of these in Fig.~\ref{fig:hashtags_before_after_election_russians}, which plots the top ten hashtags that Russian and Iranian trolls posted with substantially different rates before and after the 2016 US Presidential election.
The set of hashtags was determined by examining the relative change in posting volume before and after the election.
From the plots we make several observations.
First, we note that more general audience hashtags remain a staple of Russian trolls before the election (the relative decrease corresponds to the overall relative decrease in troll activity following the Crimea conflict).
They also use relatively innocuous/ephemeral hashtags like \#IHatePokemonGoBeacause, likely in an attempt to hide the true nature of their accounts.
That said, we also see them attaching to politically divisive hashtags like \#BlackLivesMatters around the time that Donald Trump won the Republican Presidential primaries in June 2016.
In the ramp up to the 2016 election, we see a variety of clearly political related hashtags, with \#MAGA seeing substantial peaks starting in early 2017 (higher than any peak during the 2016 Presidential campaigns).
We also see a large number of politically ephemeral hashtags attacking Obama and a campaign to push the border wall between Mexico.
In addition to these politically oriented hashtags, we again see the usage of ephemeral hashtags related to holidays.
\#SurvivalGuideToThanksgiving in late November 2016 is particularly interesting as it was heavily used for discussing how to deal with interacting with family members with wildly different view points on the recent election results.
This hashtag was exclusively used to give trolls a vector to sow discord.
When it comes to Iranian trolls, we note that, prior to the 2016 election, they share many posts with hashtags related to Hillary Clinton (see Fig.~\ref{subfig:iranians_elections_before}).
After the election they shift to posting negatively about Donald Trump (see Fig.~\ref{subfig:iranians_elections_after}).

\begin{table*}[t!]
\centering
\resizebox{0.95\textwidth}{!}{
\begin{tabular}{rl|rl}
\hline
\textbf{Topic} & \multicolumn{1}{l|}{\textbf{Terms (Russian trolls on Twitter)}} & \textbf{Topic} & \textbf{Terms (Iranian trolls on Twitter)} \\ \hline
1 & news, showbiz, photos, baltimore, local, weekend, stocks, friday, small, fatal & 1 & isis, first, american, young, siege, open, jihad, success, sydney, turkey   \\
2 & like, just, love, white, black, people, look, got, one, didn   & 2 & can, people, just, don, will, know, president, putin, like, obama  \\
3 & day, will, life, today, good, best, one, usa, god, happy  & 3 & trump, states, united, donald, racist, society, structurally, new, toonsonline, president \\
4 & can, don, people, get, know, make, will, never, want, love   & 4 & saudi, yemen, arabia, israel, war, isis, syria, oil, air, prince \\
5 & trump, obama, president, politics, will, america, media, breaking, gop, video & 5 & iran, front, press, liberty, will, iranian, irantalks, realiran, tehran, nuclear \\
6 & news, man, police, local, woman, year, old, killed, shooting, death & 6 & attack, usa, days, terrorist, cia, third, pakistan, predict, cfb, cfede \\
7 & sports, news, game, win, words, nfl, chicago, star, new, beat & 7 & israeli, israel, palestinian, palestine, gaza, killed, palestinians, children, women, year \\
8 & hillary, clinton, now, new, fbi, video, playing, russia, breaking, comey & 8 & state, fire, nation, muslim, muslims, rohingya, syrian, sets, ferguson, inferno  \\
9 & news, new, politics, state, business, health, world, says, bill, court  & 9 &  syria, isis, turkish, turkey, iraq, russian, president, video, girl, erdo  \\
10 & nyc, everything, tcot, miss, break, super, via, workout, hot, soon & 10 &  iran, saudi, isis, new, russia, war, chief, israel, arabia, peace  \\\hline
\end{tabular}
}
\caption{Terms extracted from LDA topics of tweets from Russian and Iranian trolls on Twitter.}
\label{tbl:lda_topics_agg}
\end{table*}

\begin{table}[t!]
\centering
\resizebox{\columnwidth}{!}{%
\begin{tabular}{rl}
\hline
\textbf{Topic} & \multicolumn{1}{l}{\textbf{Terms (Russian trolls on Reddit)}} \\ \hline
1 & like, also, just, sure, korea, new, crypto, tokens, north, show \\
2 & police, cops, man, officer, video, cop, cute, shooting, year, btc \\
3 & old, news, matter, black, lives, days, year, girl, iota, post\\
4 & tie, great, bitcoin, ties, now, just, hodl, buy, good, like\\
5 & media, hahaha, thank, obama, mass, rights, use, know, war, case\\
6 & man, black, cop, white, eth, cops, american, quite, recommend, years\\
7 & clinton, hillary, one, will, can, definitely, another, job, two, state\\
8 & trump, will, donald, even, well, can, yeah, true, poor, country\\
9 & like, people, don, can, just, think, time, get, want, love\\
10 & will, can, best, right, really, one, hope, now, something, good\\ \hline
\end{tabular}%
}
\caption{Terms extracted from LDA topics of posts from Russian trolls on Reddit.}
\label{tbl:lda_reddit}
\end{table}

\descr{LDA analysis.}
We also use the Latent Dirichlet Allocation (LDA) model~\cite{blei2003latent} to analyze
tweets' semantics.
We train an LDA model for each of the datasets and extract ten distinct topics with ten words,
as reported in Table~\ref{tbl:lda_topics_agg}.
While both Russian and Iranian trolls tweet about politics related topics, for Iranian trolls, this seems to be focused more on regional, and possibly even internal issues.
For example, ``iran'' itself is a common term in several of the topics, as is ``israel,'' ``saudi,'' ``yemen,'' and ``isis.''
While both sets of trolls discuss the proxy war in Syria (in which both states are involved), the Iranian trolls have topics pertaining to Russia and Putin, while the Russian trolls do not make any mention of Iran, instead focusing on more vague political topics like gun control and racism.
For Russian trolls on Reddit (see Table~\ref{tbl:lda_reddit}) we again find topics related to politics and cryptocurrencies (e.g., topic 4).

\begin{figure}[t]
\centering
\includegraphics[width=0.8\columnwidth]{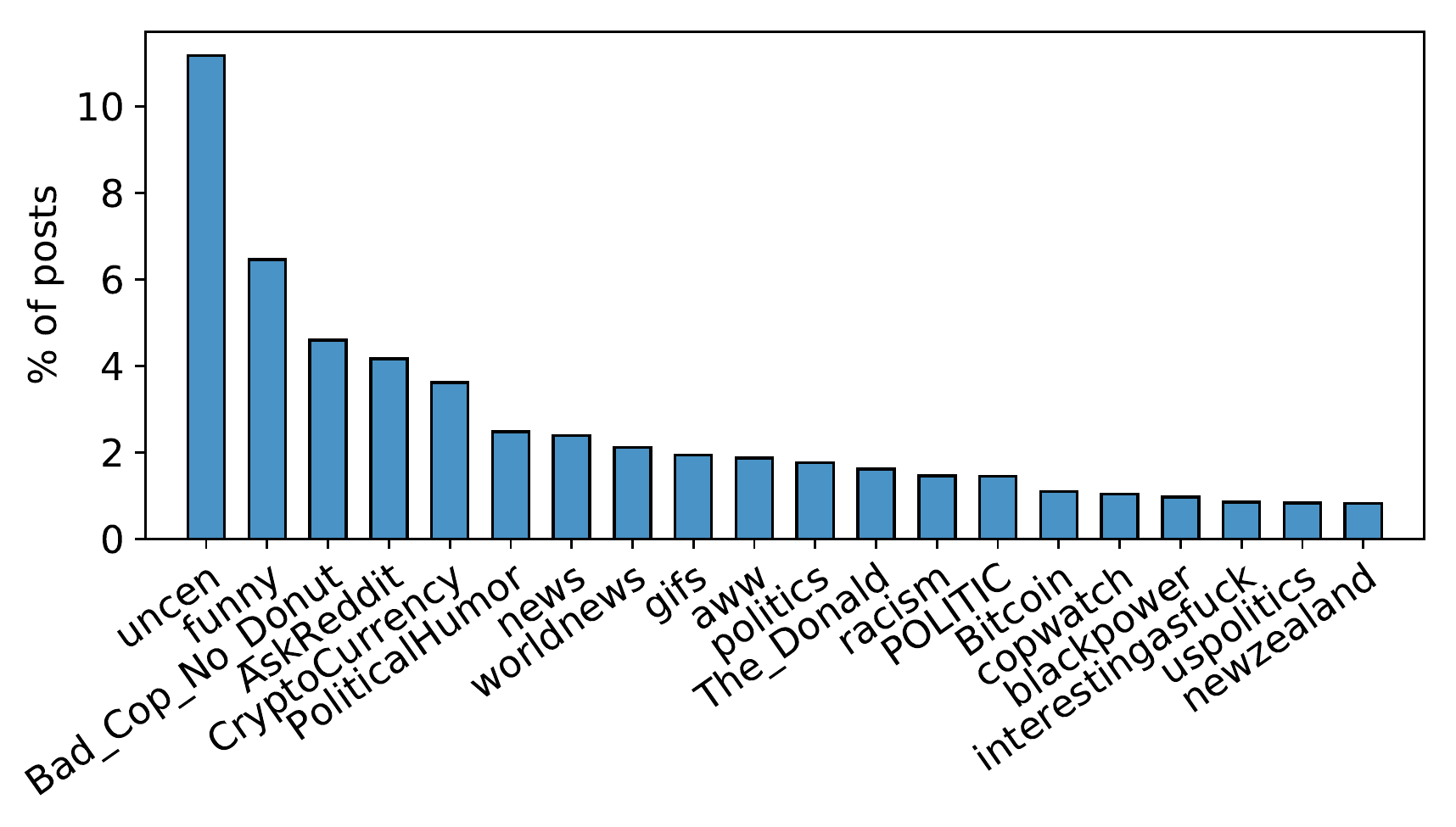}
   \caption{Top 20 subreddits that Russian trolls were active and their respective percentage of posts.}
\label{fig:top_subreddits}
\end{figure}

\descr{Subreddits.} Fig.~\ref{fig:top_subreddits} shows the top 20 subreddits that Russian trolls on Reddit exploited and their respective percentage of posts over the whole dataset.
The most popular subreddit is /r/uncen (11\% of posts), which is a subreddit created by a specific Russian troll and, via manual examination, appears to be primarily used to disseminate news articles of questionable credibility.
Other popular subreddits include general audience subreddits like /r/funny (6\%) and /r/AskReddit (4\%), likely in an attempt to obfuscate the fact that they are state-sponsored trolls in the same way that innocuous hashtags were used on Twitter.
Finally, it is worth noting that the Russian trolls were particularly active on communities related to cryptocurrencies like  /r/CryptoCurrency (3.6\%) and /r/Bitcoin (1\%) possibly attempting to influence the prices of specific cryptocurrencies.
This is particularly noteworthy considering cryptocurrencies have been reportedly used to launder money, evade capital controls, and perhaps used to evade sanctions~\cite{crypto_money_laundering,crypto_tax_evasion}.

\begin{table}[t]
\centering
\footnotesize
\resizebox{\columnwidth}{!}{%
\begin{tabular}{rrrrrl}
\hline
\textbf{\begin{tabular}[c]{@{}r@{}}Domain (Russian \\ trolls on Twitter\end{tabular}} & \multicolumn{1}{l}{\textbf{(\%)}} & \textbf{\begin{tabular}[c]{@{}r@{}}Domain(Iranian \\ trolls on Twitter)\end{tabular}} & \multicolumn{1}{l}{\textbf{(\%)}} & \textbf{\begin{tabular}[c]{@{}r@{}}Domain (Russian \\ trolls on Reddit)\end{tabular}} & \textbf{(\%)} \\ \hline
livejournal.com & \multicolumn{1}{r|}{5.4\%} & awdnews.com & \multicolumn{1}{r|}{29.3\%} & imgur.com & 27.6\% \\
riafan.ru & \multicolumn{1}{r|}{5.0\%} & dlvr.it & \multicolumn{1}{r|}{7.1\%} & blackmattersus.com & 8.3\% \\
twitter.com & \multicolumn{1}{r|}{2.5\%} & fb.me & \multicolumn{1}{r|}{4.8\%} & donotshoot.us & 3.6\% \\
ift.tt & \multicolumn{1}{r|}{1.8\%} & whatsupic.com & \multicolumn{1}{r|}{4.2\%} & reddit.com & 1.9\% \\
ria.ru & \multicolumn{1}{r|}{1.8\%} & googl.gl & \multicolumn{1}{r|}{3.9\%} & nytimes.com & 1.5\% \\
googl.gl & \multicolumn{1}{r|}{1.7\%} & realnienovosti.com & \multicolumn{1}{r|}{2.1\%} & theguardian.com & 1.4\%  \\
dlvr.it & \multicolumn{1}{r|}{1.5\%} & twitter.com & \multicolumn{1}{r|}{1.7\%} & cnn.com & 1.3\% \\
gazeta.ru & \multicolumn{1}{r|}{1.4\%} & libertyfrontpress.com & \multicolumn{1}{r|}{1.6\%} & foxnews.com & 1.2\% \\
yandex.ru & \multicolumn{1}{r|}{1.2\%} & iuvmpress.com & \multicolumn{1}{r|}{1.5\%} & youtube.com & 1.2\% \\
j.mp & \multicolumn{1}{r|}{1.1\%} & buff.ly & \multicolumn{1}{r|}{1.4\%} & washingtonpost.com & 1.2\% \\
rt.com & \multicolumn{1}{r|}{0.8\%} & 7sabah.com & \multicolumn{1}{r|}{1.3\%} & huffingntonpost.com & 1.1\% \\
nevnov.ru & \multicolumn{1}{r|}{0.7\%} & bit.ly & \multicolumn{1}{r|}{1.2\%} & photographyisnotacrime.com & 1.0\% \\
youtu.be & \multicolumn{1}{r|}{0.6\%} & documentinterdit.com & \multicolumn{1}{r|}{1.0\%} & butthis.com  & 1.0\% \\
vesti.ru & \multicolumn{1}{r|}{0.5\%} & facebook.com & \multicolumn{1}{r|}{0.8\%} & thefreethoughtproject.com  & 0.9\% \\
kievsmi.net & \multicolumn{1}{r|}{0.5\%} & al-hadath24.com & \multicolumn{1}{r|}{0.7\%} & dailymail.co.uk & 0.7\% \\
youtube.com & \multicolumn{1}{r|}{0.5\%} & jordan-times.com & \multicolumn{1}{r|}{0.7\%} & rt.com & 0.7\% \\
kiev-news.com & \multicolumn{1}{r|}{0.5\%} & iuvmonline.com & \multicolumn{1}{r|}{0.6\%} & politico.com & 0.6\% \\
inforeactor.ru & \multicolumn{1}{r|}{0.4\%} & youtu.be & \multicolumn{1}{r|}{0.6\%} & reuters.com & 0.6\% \\
lenta.ru & \multicolumn{1}{r|}{0.4\%} & alwaght.com & \multicolumn{1}{r|}{0.5\%} & youtu.be & 0.6\% \\
emaidan.com.ua & \multicolumn{1}{r|}{0.3\%} & ift.tt & \multicolumn{1}{r|}{0.5\%} & nbcnews.com & 0.6 \% \\ \hline
\end{tabular}
}
\caption{Top 20 domains included in tweets/posts from Russian and Iranian trolls on Twitter and Reddit.}
\label{tbl:top_domains}
\end{table}

\descr{URLs.}
We next analyze the URLs included in the tweets/posts.
In Table~\ref{tbl:top_domains}, we report the top 20 domains for both Russian and Iranian trolls.
Livejournal (5.4\%) is the most popular domain in the Russian trolls dataset on Twitter, likely due the Ukrainian campaign. 
Overall, we can observe the impact of the Crimean conflict, with essentially all domains posted by the Russian trolls being Russian language or Russian oriented.
One exception to Russian language sites is RT, the Russian-controlled propaganda outlet.
The Iranian trolls similarly post more ``localized'' domains, for example, jordan-times, but we also see them heavily pushing the IUVM fake news network.
When it comes to Russian trolls on Reddit, we find that they were mostly posting random images through Imgur (image-hosting site, 27.6\% of the URLs), likely in an attempt to accumulate karma score.
We also note a substantial portion of URLs to (fake) news sites linked with the Internet Research Agency like blackmattersus.com (8.3\%) and donotshootus.us (3.6\%).

\begin{table}[]
\resizebox{\columnwidth}{!}{
\begin{tabular}{@{}lrrrrrrrrr@{}}
\toprule
\multicolumn{8}{c}{\textbf{Events per community}}                                                                                                                                                                                                                                                                                                            & \multicolumn{2}{c}{\textbf{Total}}                                       \\ \midrule
\textbf{\begin{tabular}[c]{@{}c@{}}URLs\\shared by\end{tabular}} & \multicolumn{1}{c}{\textbf{\dspol}} & \multicolumn{1}{c}{\textbf{Reddit}} & \multicolumn{1}{c}{\textbf{Twitter}} & \multicolumn{1}{c}{\textbf{Gab}} & \multicolumn{1}{c}{\textbf{The\_Donald}} & \multicolumn{1}{c}{\textbf{Iran}} & \multicolumn{1}{c|}{\textbf{Russia}} & \multicolumn{1}{c|}{\textbf{Events}} & \multicolumn{1}{c}{\textbf{URLs}} \\ \hline
\textbf{Russians}                                                 & 76,155                                             & 366,319                             & 1,225,550                            & 254,016                          & 61,968                                   & 0                                 & \multicolumn{1}{r|}{151,222}         & \multicolumn{1}{r|}{2,135,230}       & 48,497                            \\
\textbf{Iranians}                                                 & 3,274                                              & 28,812                              & 232,898                              & 5,763                            & 971                                      & 19,629                            & \multicolumn{1}{r|}{0}               & \multicolumn{1}{r|}{291,347}         & 4,692                             \\
\textbf{Both}                                                     & 331                                                & 2,060                               & 85,467                               & 962                              & 283                                      & 334                               & \multicolumn{1}{r|}{565}             & \multicolumn{1}{r|}{90,002}          & 153                               \\ \bottomrule
\end{tabular}
}
\caption{Total number of events in each community for URLs shared by a) Russian trolls; b) Iranian trolls; and c) Both Russian and Iranian trolls.}
\label{tbl:hawkes}
\end{table}

\begin{figure*}[t]
\center
\subfigure[Russian trolls]{\includegraphics[width=0.33\textwidth]{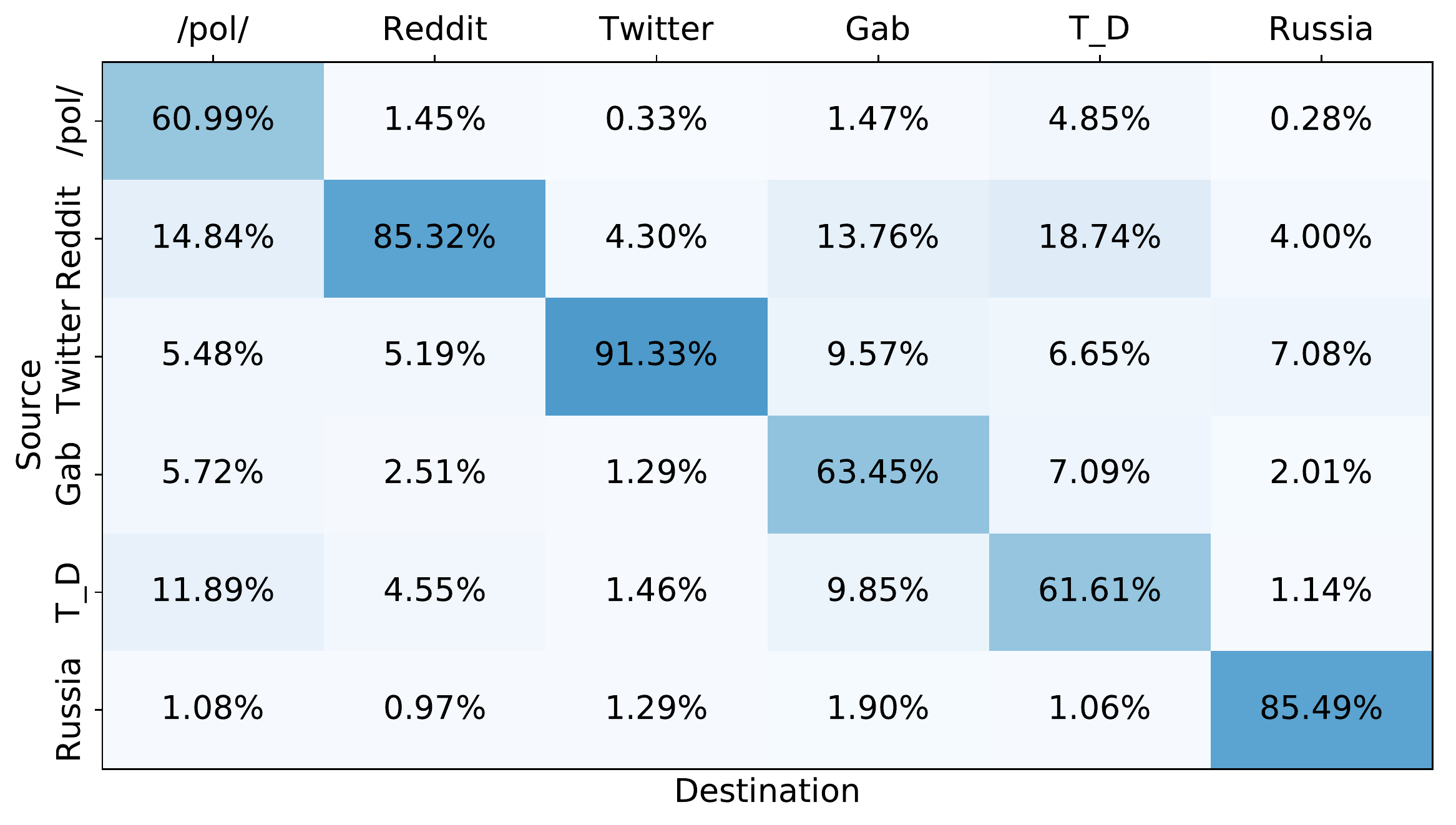}\label{subfig:raw_influence_russians}}
\subfigure[Iranian trolls]{\includegraphics[width=0.33\textwidth]{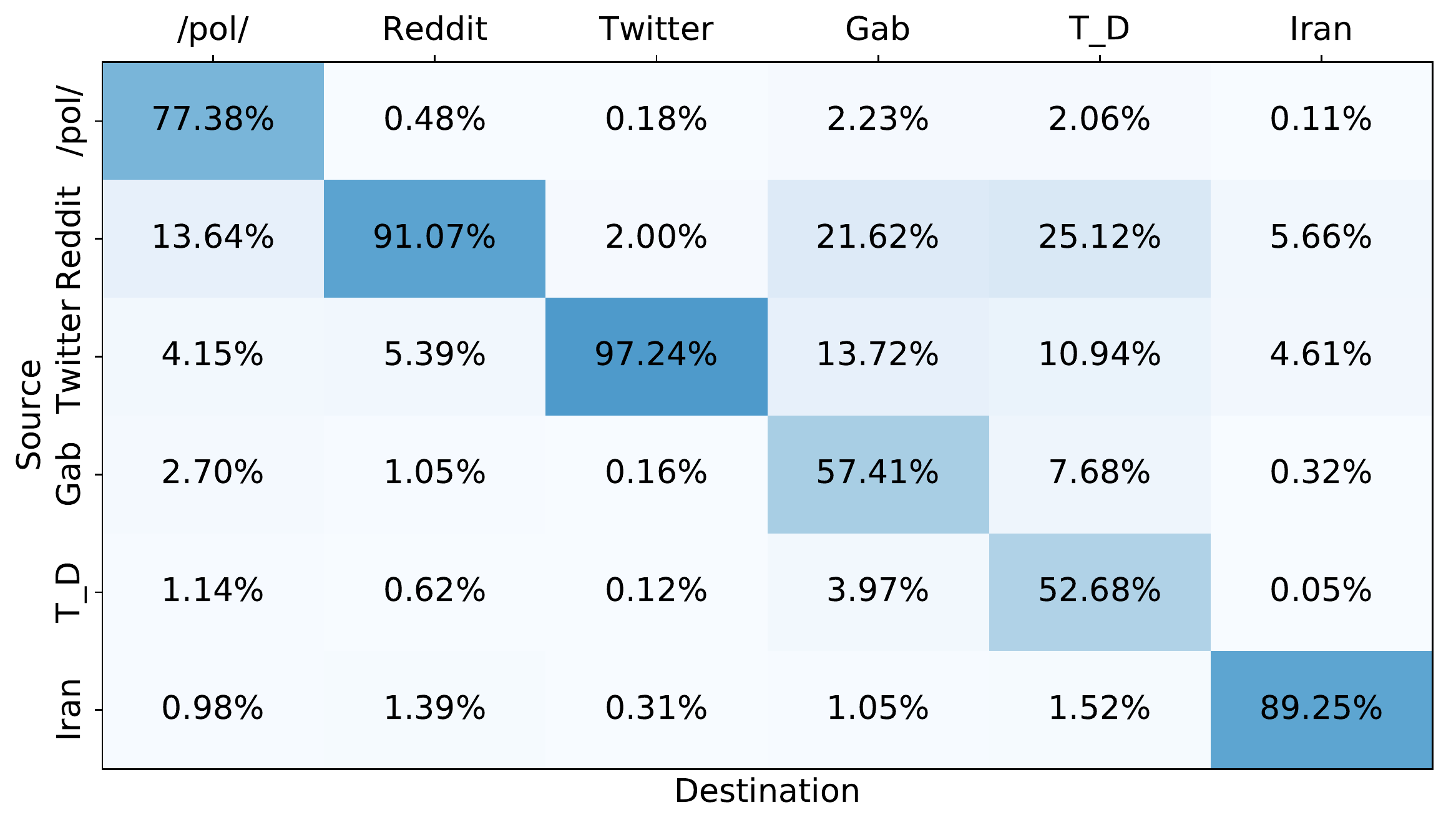}\label{subfig:raw_influence_iranians}}
\subfigure[Both]{\includegraphics[width=0.33\textwidth]{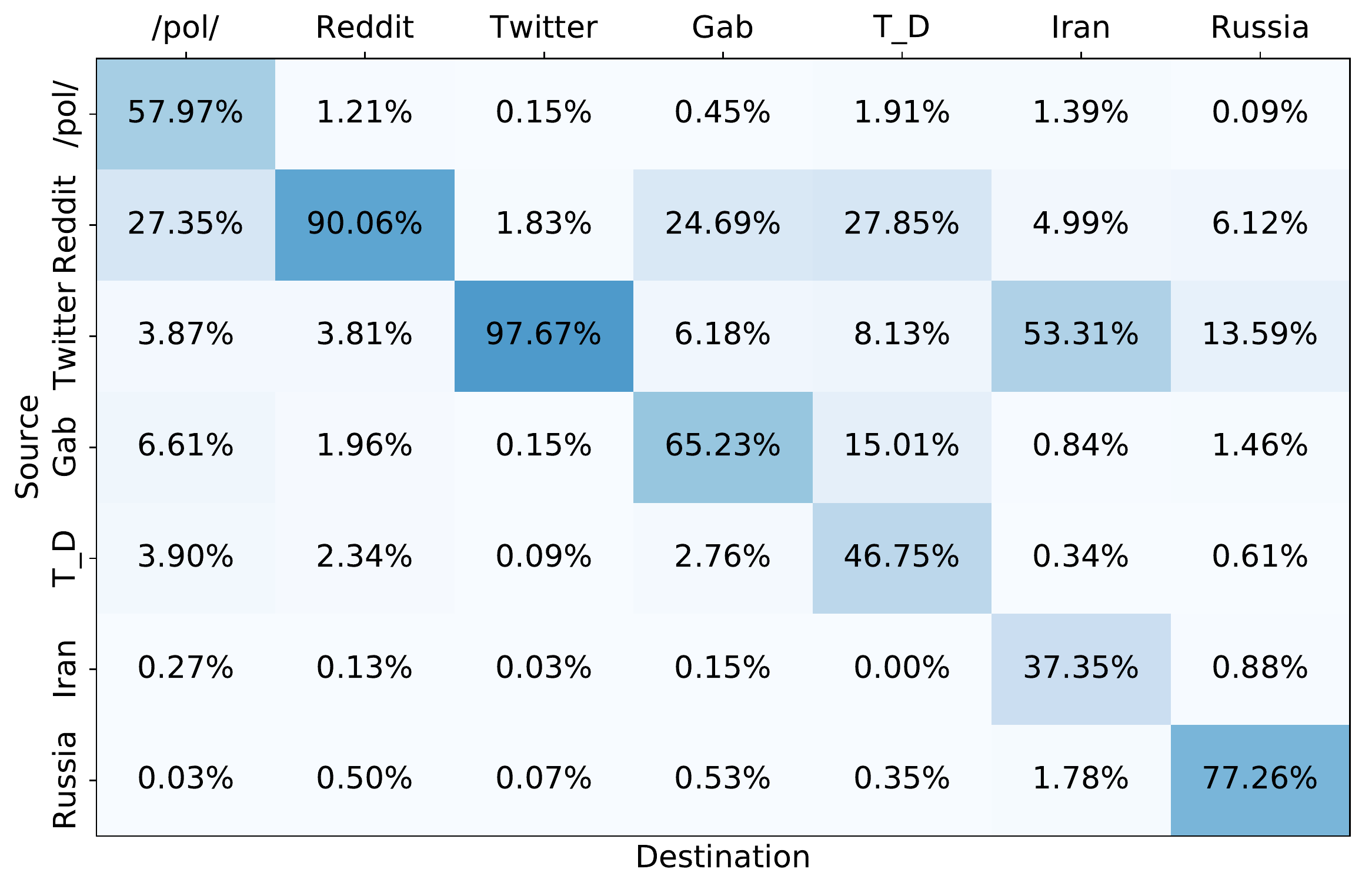}\label{subfig:raw_influence_both}}
\caption{Percent of \emph{destination} events caused by the source community to the destination community for URLs shared by a) Russian trolls; b) Iranian trolls; and c) both Russian and Iranian trolls. }
\label{fig:raw_influence}
\end{figure*}

\begin{figure*}[t]
\center
\subfigure[Russian trolls]{\includegraphics[width=0.33\textwidth]{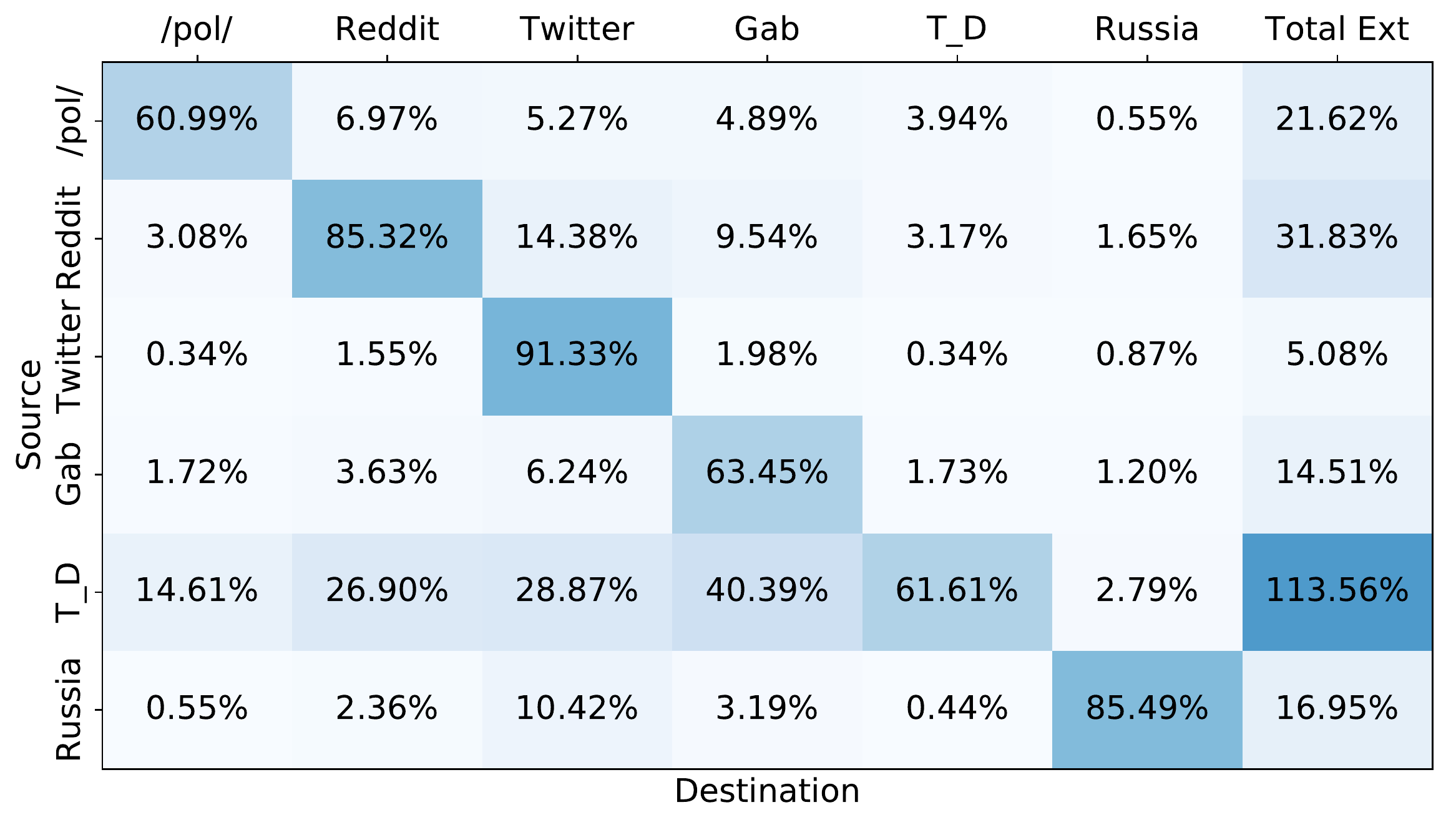}\label{subfig:norm_influence_russians}}
\subfigure[Iranian trolls]{\includegraphics[width=0.33\textwidth]{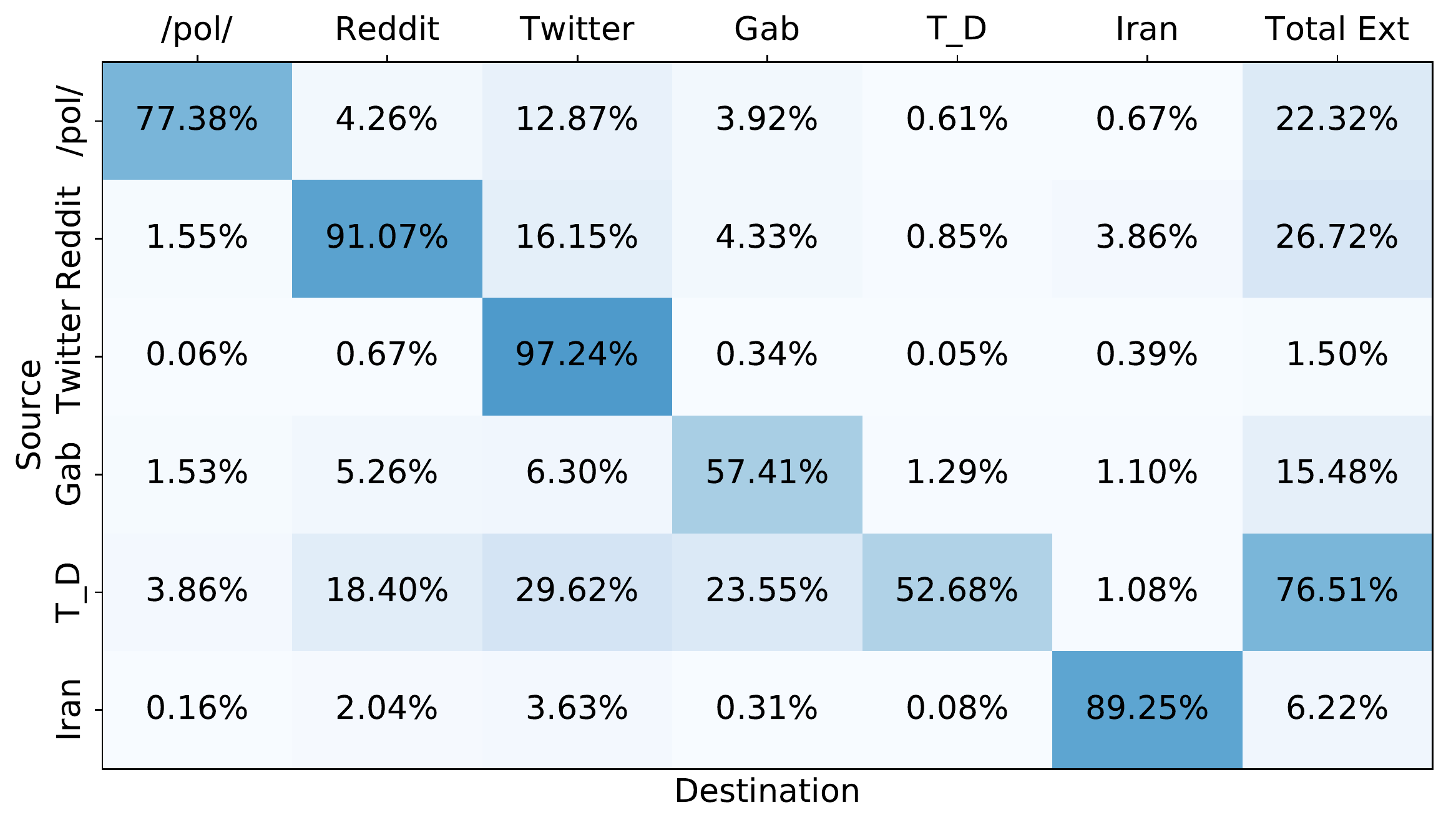}\label{subfig:norm_influence_iranians}}
\subfigure[Both]{\includegraphics[width=0.33\textwidth]{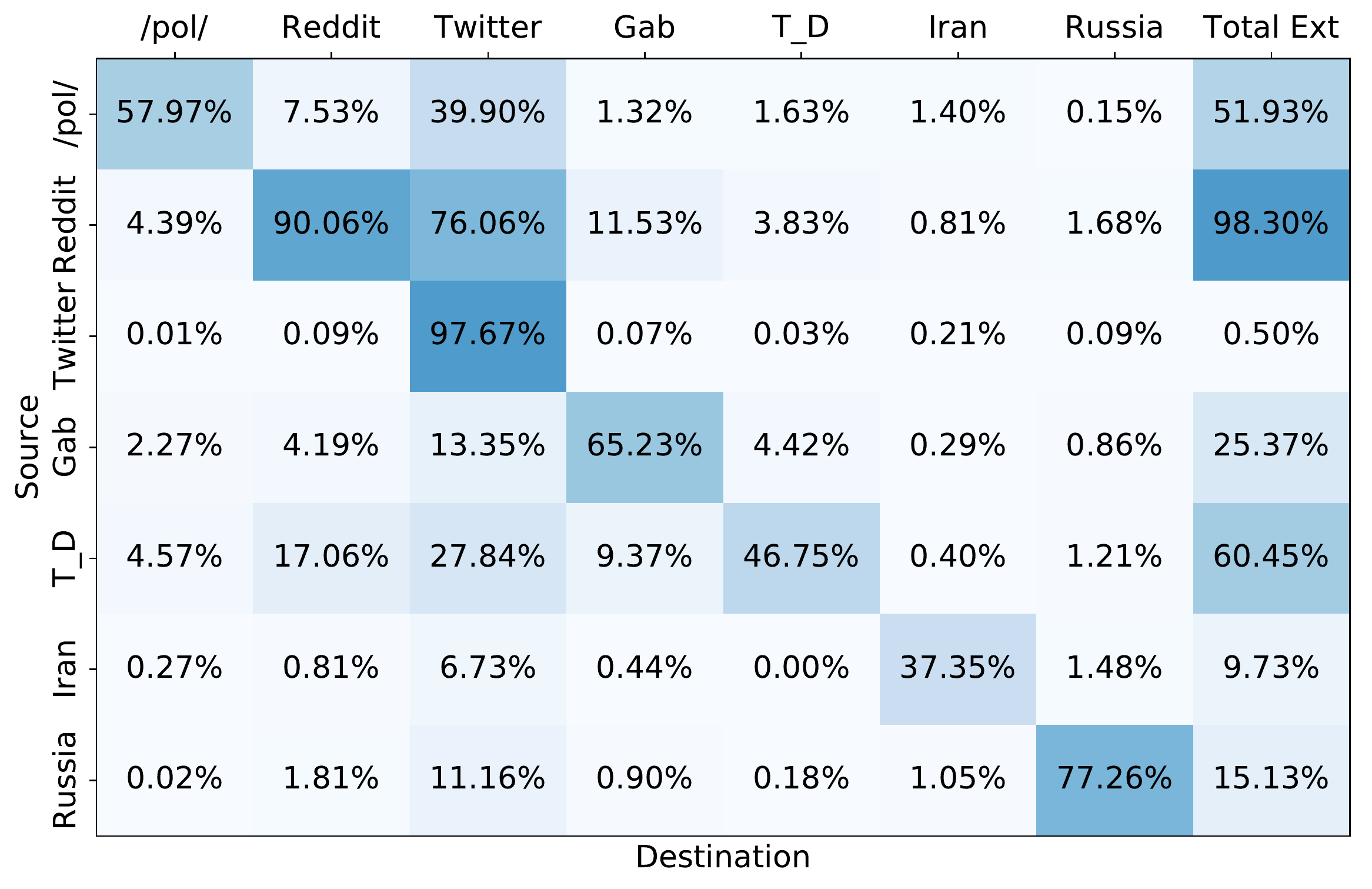}\label{subfig:norm_influence_both}}
\caption{Influence from source to destination community, normalized by the number of events in the \emph{source} community for URLs shared by a) Russian trolls; b) Iranian trolls; and c) Both Russian and Iranian trolls. We also include the total external influence of each community.}
\label{fig:norm_influence}
\end{figure*}

\section{Influence Estimation} \label{sec:influence}
Thus far, we have analyzed the behavior of Russian and Iranian trolls on Twitter and Reddit, with a special focus on how they evolved over time.
Allegedly, one of their main goals is to manipulate the opinion of other users and extend the cascade of information that they share (e.g., lure other users into posting similar content)~\cite{newsweek_manipulation}.
Therefore, we now set out to determine their impact in terms of the dissemination of information on Twitter, and on the greater Web.

To assess their influence, we look at three different groups of URLs: 1) URLs shared by Russian trolls on Twitter, 2) URLs shared by Iranian trolls on Twitter, and 3) URLs shared by both Russian \emph{and} Iranian trolls on Twitter.
We then find all posts that include any of these URLs in the following Web communities: Reddit, Twitter (from the 1\% Streaming API, with posts from confirmed Russian and Iranian trolls removed), Gab, and 4chan's Politically Incorrect board (\dspol).
For Reddit and Twitter our dataset spans January 2016 to October 2018, for \dspol it spans  July 2016 to October 2018, and for Gab it spans August 2016 to October 2018.\footnote{\textbf{NB:} the 4chan dataset made available by the authors of~\cite{zannettou2017web,zannettou2018origins} starts in late June 2016 and Gab was first launched in August 2016.}
We select these communities as previous work shows they play an important and influential role on the dissemination of news~\cite{zannettou2017web} and memes~\cite{zannettou2018origins}.

Table~\ref{tbl:hawkes} summarizes the number of events (i.e., occurrences of a given URL) for each community/group of users that we consider (Russia refers to Russian trolls on Twitter, while Iran refers to Iranian trolls on Twitter).
Note that we decouple The\_Donald from the rest of Reddit as previous work showed that it is quite efficient in pushing information in other communities~\cite{zannettou2018origins}.
From the table we make several observations: 1)~Twitter has the largest number of events in all groups of URLs mainly because it is the largest community and 2)~Gab has a considerably large number of events; more than \dspol and The\_Donald, which are bigger communities.

For each unique URL, we fit a statistical model known as Hawkes Processes~\cite{linderman2014,lindermanArxiv}, which allows us to estimate the strength of connections between each of these communities in terms of how likely an event -- the URL being posted by either trolls or normal users to a particular platform -- is to cause subsequent events in each of the groups.
We fit each Hawkes model using the methodology presented by~\cite{zannettou2018origins}.
In a nutshell, by fitting a Hawkes model we obtain all the necessary parameters that allow us to assess the root cause of each event (i.e., the community that is ``responsible'' for the creation of the event).
By aggregating the root causes for all events we are able to measure the influence and efficiency of each Web community we considered.

We demonstrate our results with two different metrics: 1)~the absolute influence, or percentage of events on the destination community caused by events on the source community and 2)~the influence relative to size, which shows the number of events caused on the destination platform as a percent of the number of events on the \emph{source} platform.
The latter can also be interpreted as a measure of how \emph{efficient} a community is in pushing URLs to other communities.

Fig.~\ref{fig:raw_influence} reports our results for the absolute influence for each group of URLs.
When looking at the influence for the URLs shared by Russian trolls on Twitter (Fig.~\ref{subfig:raw_influence_russians}), we find that Russian trolls were particularly influential to users from Gab (1.9\%), the rest of Twitter (1.29\%), and \dspol (1.08\%).
When looking at the communities that influenced the Russian trolls we find the rest of Twitter (7\%) followed by Reddit (4\%).
By looking at URLs shared by Iranian trolls on Twitter (Fig.~\ref{subfig:raw_influence_iranians}), we find that Iranian trolls were most successful in pushing URLs to The\_Donald (1.52\%), the rest of  Reddit (1.39\%), and Gab (1.05\%), somewhat ironic considering The\_Donald and Gab's zealous pro-Trump leanings and the Iranian trolls' clear anti-Trump leanings~\cite{flores2018mobilizing,zannettou2018gab}.
Similarly to Russian trolls, the Iranian trolls were most influenced by Reddit (5.6\%) and the rest of Twitter (4.6\%).
When looking at the URLs posted by both Russian and Iranian trolls we find that, overall, the Russian trolls were more influential in spreading URLs to the other Web communities with the exception of (again, somewhat ironically) \dspol.

But how do these results change when we normalize the influence with respect to the number of events that each community creates?
Fig.~\ref{fig:norm_influence} shows the influence relative to size for each pair of communities/groups of users.
For URLs shared by Russian trolls (Fig.~\ref{subfig:norm_influence_russians}) we find that Russian trolls were particularly efficient in spreading the URLs to Twitter (10.4\%)---which is not a surprise, given that the accounts operate directly on this platform---and Gab (3.19\%).
For the URLs shared by Iranian trolls, we again observe that were most efficient in pushing the URLs to Twitter (3.6\%), and the rest of Reddit (2.04\%).
Also, it is worth noting that in both groups of URLs The\_Donald had the highest external influence to the other platforms.
This highlights that The\_Donald is an impactful actor in the information ecosystem and is quite possibly exploited by trolls as a vector to push specific information to other communities.
Finally, when looking at the URLs shared by both Russian and Iranian trolls, we find that Russian trolls were more efficient (greater impact relative to the number of URLs posted) at spreading URLs in all the communities with the exception of \dspol, where Iranians were more efficient.

\section{Discussion \& Conclusion} \label{sec:conclusion}
In this paper, we analyzed the behavior and evolution of Russian and Iranian trolls on Twitter and Reddit during the course of several years.
We shed light to the target campaigns of each group of trolls, we examined how their behavior evolved over time, and what content they disseminated.
Furthermore, we find some interesting differences between the trolls depending on their origin and the platform from which they operate.
For instance, for the latter, we find discussions related to cryptocurrencies only on Reddit by Russian trolls, while for the former we find that Russian trolls were pro-Trump and Iranian trolls anti-Trump.
Also, we quantify the influence that these state-sponsored trolls had on several mainstream and alternative Web communities (Twitter, Reddit, \dspol, and Gab), showing that Russian trolls were more efficient and influential in spreading URLs on other Web communities than Iranian trolls, with the exception of \dspol.
In addition, we make our source code publicly available~\cite{code}, which helps in reproducing our results and it is an important step towards understanding other types of state-sponsored troll accounts on Twitter.

Our findings have serious implications for society at large.
First, our analysis shows that while troll accounts use peculiar tactics and talking points to further their agendas, these are not completely disjoint from regular users, and therefore developing automated systems to identify and block such accounts remains an open challenge.
Second, our results also indicate that automated systems to detect trolls are likely to be difficult to realize: trolls change their behavior over time, and thus even a classifier that works perfectly on one campaign might not catch future campaigns.
Third, and perhaps most worrying, we find that state-sponsored trolls have a meaningful amount of influence on fringe communities like The\_Donald, 4chan's \dspol, and Gab, and that the topics pushed by the trolls resonate strongly with these communities.
This might be due to users on these communities that sympathize with the views  the trolls aim to share (i.e., ``useful idiots'') or to unidentified state-sponsored actors on these communities.
In either case, considering recent tragic events like the Tree of Life Synagogue shootings, perpetrated by a Gab user seemingly influenced by content posted there, the potential for mass societal upheaval cannot be overstated.
Because of this, we implore the research community, as well as governments and non-government organizations to expend whatever resources are at their disposal to develop technology and policy to address this new, and effective, form of digital warfare.

\descr{Acknowledgments.} This project has received funding from the European Union's Horizon 2020 Research and Innovation program under the Marie Sk\l{}odowska-Curie ENCASE project (Grant Agreement No. 691025).
This work reflects only the authors' views; the Agency and the Commission are not responsible for any use that may be made of the information it contains.

\small

\bibliographystyle{abbrv}

\end{document}